# Buoyancy induced motion of a Newtonian drop in Elastoviscoplastic materials

G. Esposito, Y. Dimakopoulos, and J. Tsamopoulos[1]

*Department of Chemical Engineering, University of Patras, Patras 26504, Greece*

We investigate theoretically the buoyancy-driven motion of a viscous drop in a yield-stress material, incorporating elastic effects represented by the Saramito-Herschel-Bulkley constitutive equation. We solve the governing equations using an open-source finite volume solver and utilizing the volume of fluid technique to accurately capture the interface between the two fluids. To validate our numerical approach, we compare our results with data from previous experimental and numerical studies. We find quantitative agreement in terms of terminal velocities and drop shapes, affirming the accuracy of our model and its numerical solution. Notably, we observe that incorporating elastic effects into the modelling of the continuous phase is essential for predicting phenomena reported in experiments, such as the inversion of the flow field behind the sedimenting drop (i.e., the negative wake) or the formation of a teardrop shape. Due to the elastoviscoplastic nature of the continuous phase, we observe that small drops remain entrapped because the buoyancy force is insufficient to fluidize the surrounding material. We investigate entrapment conditions using two different protocols, which yield different outcomes due to the interplay between capillarity and elasto-plasticity. Finally, we conduct an extensive parametric analysis to evaluate the impact of rheological parameters (yield stress, elastic modulus, and interfacial tension) on the dynamics of sedimentation.

[1] tsamo@chemeng.upatras.gr

## 1. Introduction

Viscoplastic fluids, commonly known as Yield-Stress Materials (YSMs), possess a fascinating characteristic: they flow only when subjected to stress exceeding a specific threshold known as yield stress; otherwise, they behave like solids. The interest in these fluids is warranted by their widespread presence in various aspects of daily life. Examples include products from the pharmaceutical and food industries, crude oil, and materials to extract it, and formulations in the construction sector, such as concrete, paint, and plaster. Moreover, YSMs often contain a secondary phase - solid, gaseous, or liquid - due to process conditions (e.g., air entrapment during mixing) or inherent material structure (e.g., emulsions). The presence of this secondary phase can be either beneficial, as in the production of flavorful gelato, or undesirable, as in the manufacturing of pharmaceuticals requiring high purity. The physical mechanisms underlying the plastic behavior observed in YSMs have been the subject of extensive study, as discussed in detail by Bonn et al. [1]. In essence, this behavior may arise from repulsive interactions dominating among material constituents (e.g., in suspensions, emulsions, foams, and granular suspensions) or from attractive chemical bonds within the network forming the material's microstructure (e.g., colloidal gels).

Historically, viscoplastic materials have been modelled using algebraic constitutive equations that relate the rate of deformation ($\tilde{\dot{\gamma}}$) to the extra stress $\tilde{\tau}$. The first example of such models dates back to 1926, when Eugene Bingham proposed his seminal one-dimensional constitutive equation [2], which provides a dual description of the material as a rigid solid (when the applied stress is lower than the yield stress $\tilde{\sigma}_y$) or as a Newtonian fluid with viscosity $\tilde{\eta}$ (when the yield stress is overcome):

$$\tilde{\dot{\gamma}} = \begin{cases} 0, & if\ \tilde{\tau} \leq \tilde{\sigma}_y \\ \dfrac{\tilde{\tau} - \tilde{\sigma}_y}{\tilde{\eta}}, & if\ \tilde{\tau} > \tilde{\sigma}_y \end{cases} \tag{1}$$

Subsequently, several modifications have been proposed to enrich the Bingham model. One of the most widely adopted is the Herschel-Bulkley model [3], where a flow-dependent viscosity $\tilde{k}\tilde{\dot{\gamma}}^{n-1}$ is defined to include shear-thinning effects:

$$\tilde{\dot{\gamma}} = \begin{cases} 0, & if\ \tilde{\tau} \leq \tilde{\sigma}_y \\ \left(\dfrac{\tilde{\tau} - \tilde{\sigma}_y}{\tilde{k}}\right)^{\frac{1}{n}} & if\ \tilde{\tau} > \tilde{\sigma}_y \end{cases} \tag{2}$$

The simplicity of models like the Bingham and Herschel-Bulkley has contributed to their widespread use. However, it is important to note that they suffer from both mathematical and physical limitations. Primarily, both models do not represent adequately the behaviour of YSMs below the yield stress, either from a physical perspective, since often such materials exhibit elastic response [4], or from a mathematical point of view, due to the fact that the stress field in the unyielded regions is unspecified; hence, any velocity and stress field satisfying the mass and momentum balance with stress below the yield condition is in principle an acceptable solution. To bypass the second limitation, two main approaches have been proposed. A first approach consists in the regularization of the constitutive model by correlating the rate of deformation to the applied stress through a continuous function representing a plastic viscosity [5], [6]. A more formal and systematic approach, although algorithmically more complex and computationally more expensive, is represented by the Augmented Lagrangian Method [7], which has been recently improved, stabilized and used for the simulations of complex flows involving viscoplastic materials [8], [9].

With the goal to address such limitations, a crucial contribution was provided by J. Oldroyd in 1947 [10]. He proposed a tensorial generalization of the Bingham model, where the unyielded regions are modelled as a Hookean elastic solid with elastic modulus $\tilde{G}$. Consequently, the extra stress in the unyielded regions is a linear function of the deformation, $\tilde{\boldsymbol{\gamma}}$, while in the yielded regions it remained a linear function of the instantaneous rate-of-deformation, $\dot{\tilde{\boldsymbol{\gamma}}}$:

$$\tilde{\boldsymbol{\tau}} = \begin{cases} \tilde{G}\tilde{\boldsymbol{\gamma}}, & if\ |\tilde{\boldsymbol{\tau}}_{\boldsymbol{d}}| < \tilde{\sigma}_y \\ \tilde{\sigma}_y + \tilde{\eta}\dot{\tilde{\boldsymbol{\gamma}}}, & if\ |\tilde{\boldsymbol{\tau}}_{\boldsymbol{d}}| > \tilde{\sigma}_y \end{cases} \tag{3}$$

Here, the term $\tilde{\boldsymbol{\tau}}_{\boldsymbol{d}}$ represents the deviatoric part of the extra stress tensor $\tilde{\boldsymbol{\tau}}_{\boldsymbol{d}}$, defined as:

$$\tilde{\boldsymbol{\tau}}_{\boldsymbol{d}} = \tilde{\boldsymbol{\tau}} - \frac{tr(\tilde{\boldsymbol{\tau}})}{3}\boldsymbol{I} \tag{4}$$

With $|\tilde{\boldsymbol{\tau}}_{\boldsymbol{d}}|$ its magnitude, defined as:

$$|\tilde{\boldsymbol{\tau}}_{\boldsymbol{d}}| = \sqrt{\frac{1}{2}\tilde{\boldsymbol{\tau}}_{\boldsymbol{d}}:\tilde{\boldsymbol{\tau}}_{\boldsymbol{d}}} \tag{5}$$

This modification allows a more adequate description of the material in the unyielded regions and represents a significant improvement from a mathematical perspective, since the stress field is now specified before and after yielding. On the other hand, this model still predicts a non-physical discontinuity in the stress for $|\tilde{\boldsymbol{\tau}}_{\boldsymbol{d}}| = \tilde{\sigma}_y$, because in the first branch the stress is proportional to the strain, while in the second branch it is proportional to the strain rate, and the transition occurs at a non-null critical strain rate. Hence, it does not represent a fully acceptable solution to the fundamental problem of modelling YSMs. Extensive reviews, including an interesting historical perspective on the rheological models developed for YSMs, can be found in Mitsoulis & Tsamopoulos [11] and Frigaard [12].

The necessity to extend and improve the modelling of YSMs is not only driven by the inability of the aforementioned models to properly describe a continuous transition from solid to fluid behaviour, but also by recent experimental observations showing that such materials often manifest an elastic response both before and after yielding. De Cagny et. al [13] performed a systematic experimental analysis on three different YSMs, measuring non-zero values of the two normal stress differences both above and below the shear yield stress. Holenberg et al. [14] analyzed experimentally the sedimentation of a spherical rigid particle in a Carbopol solution, a very often used and transparent yield stress material. The occurrence of a negative wake behind the falling sphere, i.e., an inversion in the direction of the flow field at the trail of the falling object, is reported by means of Particle Image Velocimetry (PIV) measurements, with consequent breakage of the fore-aft symmetry of the flow field at negligible inertia. Such experimental results contradict the theoretical predictions concerning the creeping flow of falling spheres in viscoplastic fluids when elastic effects are not taken into account [15], but have been successfully reproduced by means of numerical simulations when elasticity is included in the modelling of the material [16]. Similar findings have been reported for the rise of air bubbles in Carbopol by Mougin, Magnin and Piau, again employing PIV measurements [17]. Previously, such phenomenon was observed only in pure viscoelastic solutions with no measurable yield stress [18], [19]. Dubash and Frigaard [20] studied the translation and stopping of air bubbles in Carbopol solutions, reporting pronounced inverted teardrop shapes for small bubbles rising in Carbopol at moderate concentrations. Such shapes are not computationally reproducible through standard viscoplastic models (Bingham, Herschel-Bulkley), regardless of the numerical method employed to deal with the solid-liquid transition [21], [22]. Recently, Lopez et al. [23] performed a similar study, concerning the rising of an air bubble in Carbopol solutions at different concentrations. Depending on their volume, the bubbles are reported to acquire different shapes, ranging from the typical oblate shapes observed in Newtonian solutions to the more elongated inverted teardrop ones, reported in polymeric fluids with strong elasticity. The proof that such shapes are not the result of uncontrolled injection conditions or confinement effects is provided by the experimental study of Pourzahedi, Zare and Frigaard [24], where the authors show that the inverted teardrop shape can only result from the elasticity of the material.

A breakthrough contribution in the modelling of YSM is represented by the constitutive equation proposed by Saramito in 2007 [25]. This model includes a viscoelastic response of the material both before and after yielding, with a continuous relation between the polymeric stress $(\tilde{\boldsymbol{\tau}}_{\boldsymbol{p}})$ and the rate of deformation that eliminates any requirement for the regularization of the constitutive equation:

$$\frac{1}{\tilde{G}}\overset{\nabla}{\tilde{\boldsymbol{\tau}}}_{\boldsymbol{p}} + max\left(0, \frac{|\tilde{\boldsymbol{\tau}}_{\boldsymbol{p,d}}| - \tilde{\sigma}_y}{\tilde{\eta}|\tilde{\boldsymbol{\tau}}_{\boldsymbol{p,d}}|}\right)\tilde{\boldsymbol{\tau}}_{\boldsymbol{p}} = \tilde{\dot{\boldsymbol{\gamma}}} \tag{6}$$

The max term in **Eq. (6)** incorporates the von Mises criterion [26] and distinguishes the yielded regions ($|\tilde{\boldsymbol{\tau}}_{\boldsymbol{p,d}}| > \tilde{\sigma}_y$), where the material behaves as a viscoelastic liquid, from the unyielded ones ($|\tilde{\boldsymbol{\tau}}_{\boldsymbol{p,d}}| < \tilde{\sigma}_y$), where a Kelvin-Voigt viscoelastic solid behaviour is recovered. As a consequence, the locus of points where $|\tilde{\boldsymbol{\tau}}_{\boldsymbol{p,d}}| = \tilde{\sigma}_y$ represents the yield surface. Furthermore, the usage of the upper convected derivative ensures the frame invariance of the constitutive equation, and the adequacy for the kinematic description of such materials when they undergo large deformations. Briefly, the material behaves as a neo-Hookean hyperelastic solid before yielding and as an Oldroyd-B like viscoelastic fluid after yielding. The transition from solid-like to fluid-like regions is governed by the von Mises criterion [26]. A modification of the model above was proposed two years later by Saramito [27]:

$$\frac{1}{\tilde{G}}\overset{\nabla}{\tilde{\boldsymbol{\tau}}}_{\boldsymbol{p}} + max\left(0, \frac{|\tilde{\boldsymbol{\tau}}_{\boldsymbol{p,d}}| - \tilde{\sigma}_y}{\tilde{k}|\tilde{\boldsymbol{\tau}}_{\boldsymbol{p,d}}|^n}\right)^{\frac{1}{n}}\tilde{\boldsymbol{\tau}}_{\boldsymbol{p}} = \tilde{\dot{\boldsymbol{\gamma}}} \tag{7}$$

This new version of the model introduces two important improvements. Primarily, the extensional viscosity in uniaxial extension experiments is predicted to be finite, regardless of the applied extension rate. Secondarily, shear thinning effects are included and modulated by the shear-thinning parameter $n$. Such modifications make this model particularly adequate for the simulations of complex flows of yield stress materials. A plethora of numerical studies concerning the rising of air bubbles in Newtonian [28], viscoplastic [21] and viscoelastic [29] fluids are available in the literature, but only recently this problem has been revisited by Moschopoulos et al. [30], employing the new constitutive equation proposed by Saramito. The authors used the newly proposed, stabilized finite element formulation for flows with elasticity [31], [32], [33] to study the rising of an air bubble in an unbounded domain filled with an elastoviscoplastic fluid, reproducing with satisfactory agreement the experimental findings of Lopez et al. [23] and Pourzahedi, Zare, Frigaard [24].

On the other hand, although it represents an interesting problem both from a fundamental and a technological perspective, the gravity-induced motion of Newtonian viscous drops in YSMs has received much less attention. Experimental studies for the settling of viscous drops in Carbopol [34]–[36] report that, depending on the Carbopol concentration and on the size of the drop, a steady teardrop shape can be established after a long transient period, suggesting the crucial role of elastic effects. Previous numerical simulations concerning the sedimentation of Newtonian drops in viscoplastic materials failed to reproduce such teardrop shapes and always predict a fore-aft symmetry in both the flow streamlines and the yield surface when inertia is negligible [37]. To our current knowledge, this is the first theoretical work describing the gravity induced motion of Newtonian drops in YSMs exhibiting elastic effects. This paper is structured as follows: Section 2 outlines the problem formulation, encompassing the governing equations, the selection of rheological parameters for the constitutive equation, and the numerical method utilized for the simulations. In Section 3, we delve into the presentation and discussion of our numerical analysis. Initially, we introduce a base case aimed at replicating the experimental results documented in the literature. Then, we perform a systematic parametric analysis to assess the influence of the yield stress $\tilde{\sigma}_y$, the elastic modulus $\tilde{G}$ and the interfacial tension $\tilde{\Gamma}$. Finally, conclusions are presented and discussed in Section 4.

## 2. Problem formulation

We examine the sedimentation of a single drop of Newtonian liquid in a material exhibiting an elastoviscoplastic behaviour via numerical simulations under axisymmetric conditions. The assumption of axial symmetry is supported by the experimental results used for comparison [35] and other works related to the buoyancy-driven motion of deformable objects in low-concentration Carbopol gels under similar conditions [23]. All vectorial and tensorial quantities are indicated in bold. Dimensional quantities are indicated with a tilde. A schematic representation of the problem setup is depicted in **Fig**. 1: a spherical liquid drop of radius $\tilde{R}$ is placed along the axis of symmetry of a cylindrical tube of radius $\tilde{R}_c$ and length $\tilde{L}$, at a distance $\tilde{d}_0 = 10\tilde{R}$ from the upper boundary. The drop initially occupies a spherical domain $\Omega_1$, while the elastoviscoplastic material occupies the remaining domain $\Omega_2$. Both the height and the radius of the cylindrical vessel are chosen to be equal to $40\tilde{R}$, to avoid affecting the flow by the confinement. To ensure that the values of $\tilde{d}_0$ and $\tilde{L}$ do not affect the dynamics of sedimentation, we conducted a study on the effect of the initial distance of the drop from the upper boundary and on the axial length of the container, whose results are reported in the Appendix.

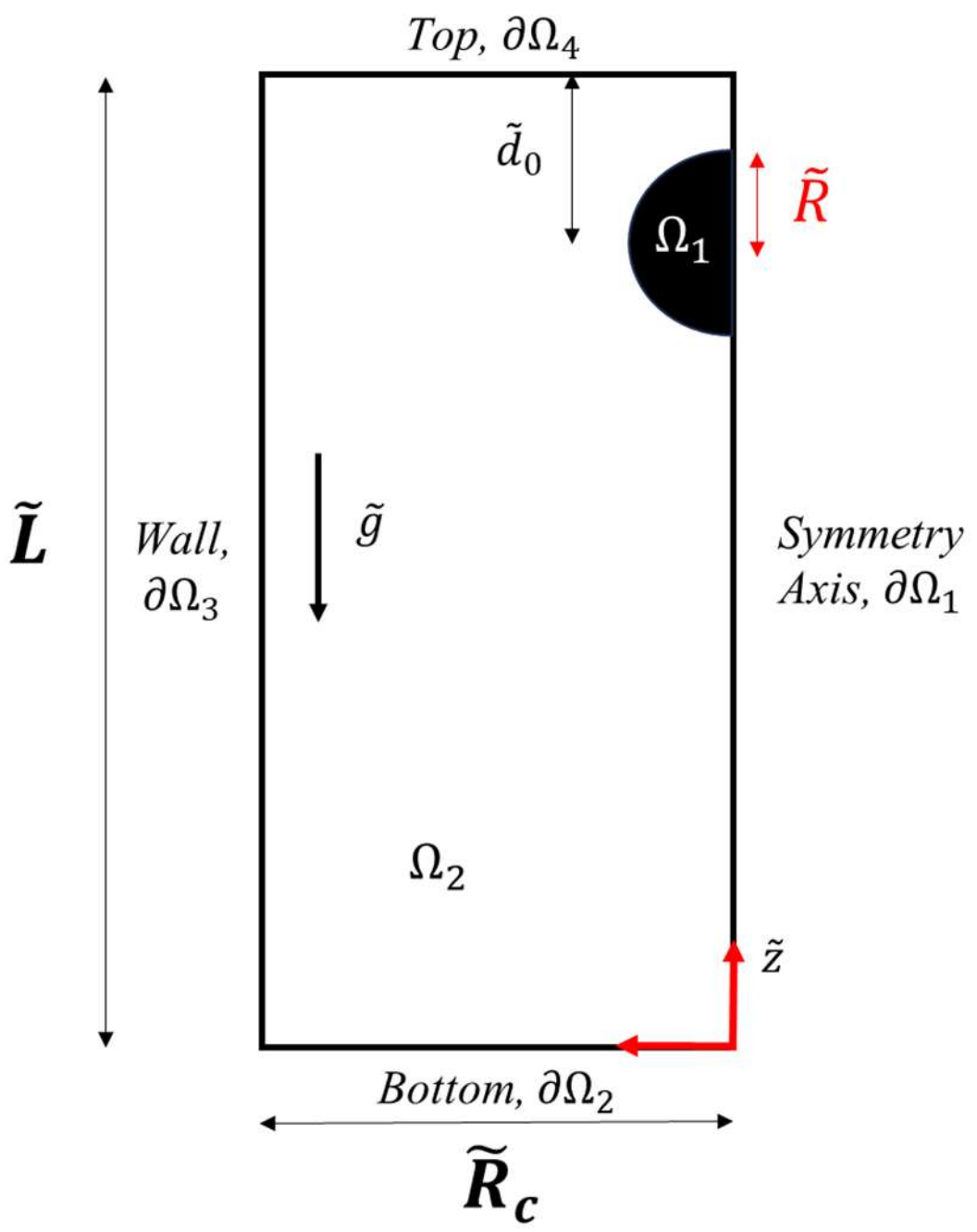

**Fig. 1**: Schematic representation of a Newtonian drop of radius $\tilde{R}$ sedimenting in a cylindrical tube of length $\tilde{L}$ and radius $\tilde{R}_c$ filled with an elastoviscoplastic material. The origin of the cylindrical coordinate system is placed on the bottom-right corner and the drop sediments along the axis of symmetry. The drop originally occupies a spherical domain $\Omega_1$, while the elastoviscoplastic fluid fills the domain $\Omega_2$.

### *2.1 Governing equations*

In this study, we assume isothermal conditions and both the EVP material (*i=2)* and the Newtonian drop *(i=1)* to be incompressible. The governing equations are the mass and momentum balance:

$$\tilde{\boldsymbol{\nabla}} \cdot \tilde{\boldsymbol{u}}_{\boldsymbol{i}} = 0 \tag{8}$$

$$\tilde{\rho}_i \left( \frac{\partial \tilde{\boldsymbol{u}}_{\boldsymbol{i}}}{\partial \tilde{t}} + \tilde{\boldsymbol{u}}_{\boldsymbol{i}} \cdot \tilde{\boldsymbol{\nabla}} \tilde{\boldsymbol{u}}_{\boldsymbol{i}} \right) = -\tilde{\boldsymbol{\nabla}} \tilde{P}_i + \tilde{\rho}_i \tilde{\boldsymbol{g}} + \tilde{\boldsymbol{\nabla}} \cdot \tilde{\boldsymbol{\tau}}_{\boldsymbol{i}} \tag{9}$$

Where $\tilde{\boldsymbol{u}}_{\boldsymbol{i}}$ is the velocity vector, $\tilde{P}_i$ is the pressure in each phase. The gravitational acceleration is indicated as $\tilde{\boldsymbol{g}}$ and the density of each phase as $\tilde{\rho}_i$. The extra stress $\tilde{\boldsymbol{\tau}}_{\boldsymbol{i}}$ is split into a Newtonian solvent contribution and a polymeric extra stress tensor $\tilde{\boldsymbol{\tau}}_{\boldsymbol{p},\boldsymbol{i}}$:

$$\tilde{\boldsymbol{\tau}}_{\boldsymbol{i}} = 2\tilde{\eta}_{s_i} \tilde{\boldsymbol{D}}_{\boldsymbol{i}} + \tilde{\boldsymbol{\tau}}_{\boldsymbol{p},\boldsymbol{i}} \tag{10}$$

where $\tilde{\boldsymbol{D}}_{\boldsymbol{i}} = \frac{1}{2}(\tilde{\boldsymbol{\nabla}} \tilde{\boldsymbol{u}}_{\boldsymbol{i}} + (\tilde{\boldsymbol{\nabla}} \tilde{\boldsymbol{u}}_{\boldsymbol{i}})^{\boldsymbol{T}})$ is the rate of strain tensor. The liquid drop is Newtonian, therefore the polymeric contribution is null, $\tilde{\boldsymbol{\tau}}_{\boldsymbol{p},\boldsymbol{1}} = \boldsymbol{0}$. In contrast, the elastoviscoplastic continuous phase is modelled by choosing an appropriate constitutive equation. We select the Saramito-Herschel-Bulkley [27] (hereafter referred as SHB) to model its rheology:

$$\frac{1}{\tilde{G}} \overset{\nabla}{\tilde{\boldsymbol{\tau}}}_{\boldsymbol{p},\boldsymbol{2}} + \max\left(0, \frac{|\tilde{\boldsymbol{\tau}}_{\boldsymbol{p},\boldsymbol{d},\boldsymbol{2}}| - \tilde{\sigma}_y}{\tilde{k}|\tilde{\boldsymbol{\tau}}_{\boldsymbol{p},\boldsymbol{d},\boldsymbol{2}}|^n}\right)^{\frac{1}{n}} \tilde{\boldsymbol{\tau}}_{\boldsymbol{p},\boldsymbol{2}} = 2\widetilde{\boldsymbol{D}_{\boldsymbol{2}}} \tag{11}$$

A schematic of the mechanical analogue of the SHB model is depicted in **Fig. 2**. Here, $\tilde{G}$ is the elastic modulus, $\tilde{k}$ is the consistency index, $n$ is the Herschel-Bulkley shear-thinning exponent and $\tilde{\sigma}_y$ is the yield stress of the material. It is important to recognize that none of the parameters of the SHB model is directly associated with a polymeric viscosity. It is possible to introduce such characteristic polymeric viscosity combining the consistency index $\tilde{k}$ and a characteristic time $\tilde{t}$, as $\tilde{\eta}_{p,2} = \frac{\tilde{k}}{\tilde{t}^{n-1}}$. This characteristic time comes from the ratio between the characteristic length and the characteristic velocity employed in this work, thus $\tilde{t} = \sqrt{\frac{R}{g}}$. The upper convected derivative (UCD) of the polymeric extra stress tensor is given by:

$$\overset{\nabla}{\tilde{\boldsymbol{\tau}}}_{\boldsymbol{p},2} = \frac{\partial \tilde{\boldsymbol{\tau}}_{\boldsymbol{p},2}}{\partial \tilde{t}} + \tilde{\boldsymbol{u}} \cdot \tilde{\boldsymbol{\nabla}} \tilde{\boldsymbol{\tau}}_{\boldsymbol{p},2} - \left(\tilde{\boldsymbol{\nabla}} \tilde{\boldsymbol{u}}\right)^{\boldsymbol{T}} \cdot \tilde{\boldsymbol{\tau}}_{\boldsymbol{p},2} - \tilde{\boldsymbol{\tau}}_{\boldsymbol{p},2} \cdot \tilde{\boldsymbol{\nabla}} \tilde{\boldsymbol{u}} \tag{12}$$

In this study, given the coexistence of two phases, an accurate method for localizing the interface is imperative. We adopt the Volume of Fluid (VOF) interface capturing technique for its numerical approximation. This approach entails a single-fluid formulation, where the local density, viscosity, velocity, and extra stress fields are determined as weighted averages of the respective properties of the two phases. The weighting factor is the indicator function $\phi$, physically representing the volumetric fraction of one of the two phases in the computational cell. The value of $\phi$ is bounded between 0 and 1. In particular, we have chosen $\phi = 1$ when the computational cell lies entirely in the domain containing the EVP material, $\phi = 0$ when it is completely occupied by the Newtonian drop, and it varies continuously in $0 < \phi < 1$ when a computational cell contains both phases and, consequently, the interface. The colour function is advected according to a scalar transport equation:

$$\frac{\partial \phi}{\partial \tilde{t}} + \tilde{\boldsymbol{u}}_{\boldsymbol{i}} \cdot \tilde{\boldsymbol{\nabla}} \phi = 0 \tag{13}$$

The density and viscosity of the system result from the weighted averages of the corresponding values in each phase:

$$\tilde{\rho} \equiv \tilde{\rho}(\phi) = \phi \tilde{\rho}_2 + (1 - \phi) \tilde{\rho}_1 \tag{14}$$

$$\tilde{\eta} \equiv \tilde{\eta}(\phi) = \phi \tilde{\eta}_2 + (1 - \phi) \tilde{\eta}_1 \tag{15}$$

While the Newtonian drop has a single constant viscosity ($\tilde{\eta}_{s,1}$), in the EVP material the sum of the solvent viscosity $\tilde{\eta}_{s,2}$ and the characteristic polymeric viscosity $\tilde{\eta}_{p,2}$ gives the total characteristic viscosity $\tilde{\eta}_2 = \tilde{\eta}_{s,2} + \tilde{\eta}_{p,2} = \tilde{\eta}_{s,2} + \frac{\tilde{k}}{\tilde{t}^{n-1}}$. The effect of capillary forces is then introduced in the single momentum balance holding in both phases, through an additional body force $\tilde{\boldsymbol{f}}_\sigma = \tilde{\Gamma} \delta \tilde{\kappa} \boldsymbol{n}$, following the "Continuum Surface Force (CSF)" method proposed by Brackbill [38]; see Eq. (23) below. The parameter $\tilde{\Gamma}$ represents the interfacial tension, $\delta$ is the Dirac distribution that activates this term only at the interface between the two fluids, $\boldsymbol{n}$ the normal vector to the fluid-fluid interface pointing toward the continuous phase, $\tilde{\kappa} = -\tilde{\nabla}_s \cdot \boldsymbol{n}$ the interface curvature, and $\tilde{\nabla}_s$ the surface gradient operator. The interfacial tension is assumed to be constant and spatially independent. To complete the problem, we impose appropriate boundary conditions on the boundaries $\partial\Omega_i$. The side wall, $\partial\Omega_3$, and the bottom, $\partial\Omega_2$, are static, and a no-slip condition is applied for the velocity vector. The top of the domain, $\partial\Omega_4$, is an open boundary where a fixed atmospheric pressure is imposed and the velocity vector satisfies a zero-gradient condition [39]. Symmetry conditions are applied on the boundary $\partial\Omega_1$. Since we are interested in the solution of the transient problem, an initial condition is required for the velocity field and for the indicator function $\phi$. Furthermore, the time-dependent character of the constitutive equation of the EVP material requires an appropriate initial condition as well. Initially, both fluids are stationary, and the EVP material is originally stress-free ($\tilde{\boldsymbol{\tau}}_{\boldsymbol{i}}|_{t=0} = \mathbf{0}$ and $\tilde{\boldsymbol{u}}_{\boldsymbol{i}}|_{t=0} = \mathbf{0}$). The initial distribution of the indicator function $\phi$ reflects that the domain $\Omega_1$ is spherical.

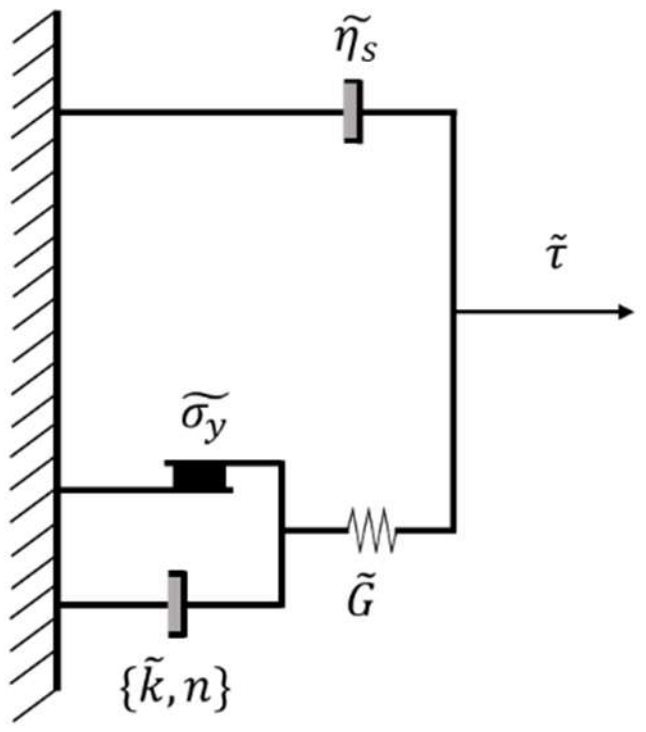

**Fig. 2:** Mechanical analogue of the SHB model. The elastic behaviour is regulated by the spring with elastic modulus $\tilde{G}$, while plastic effects are associated with the solid friction element with threshold $\tilde{\sigma}_y$. Notice that for $\tilde{G} \to \infty$, the inelastic Herschel-Bulkley model is recovered, while for $\tilde{\sigma}_y \to 0$ and $n \to 1$ (i.e., null yield stress and constant shear viscosity), the Oldroyd-B model is obtained.

### *2.2 Scaling and dimensionless quantities*

We solve the aforementioned set of equations in their dimensionless form, after defining the characteristic values of all variables (indicated by *). We scale the lengths with the effective radius of the drop, $\tilde{R}$, corresponding to the radius of the initially spherical drop. The velocity is scaled by balancing inertial terms with buoyancy, and we calculate the terminal velocity of the drop as part of the solution. The characteristic values for stresses and pressure are obtained through a balance with the buoyancy term:

$$\tilde{L}^* \to \tilde{R}; \quad \tilde{\boldsymbol{u}}^* \to \sqrt{\tilde{g}\tilde{R}}; \quad \tilde{t}^* = \sqrt{\frac{\tilde{R}}{\tilde{g}}}; \quad \tilde{\boldsymbol{\tau}}^* \to \Delta\tilde{\rho}\tilde{g}\tilde{R}; \quad \tilde{P}^* \to \Delta\tilde{\rho}\tilde{g}\tilde{R} \tag{16}$$

Here the density difference is defined as $\Delta\tilde{\rho} = \tilde{\rho}_1 - \tilde{\rho}_2$, which is positive in this study. As a consequence, the corresponding characteristic polymeric viscosity is defined as $\tilde{\eta}_{p,2} = \tilde{k}\left(\sqrt{\frac{\tilde{g}}{\tilde{R}}}\right)^{n-1}$.

The dimensionless numbers arising in the formulation of the problem, together with their definition and physical meaning, are summarized in **Table I**:

**Table I:** Dimensionless numbers and their physical meaning

| Dimensionless number | Definition | Ratio of |
|---|---|---|
| Density ratio, $\rho^{\circ}$ | $\dfrac{\tilde{\rho}_1}{\tilde{\rho}_2}$ | Density of the drop and density of the EVP material |
| Viscosity ratio, $\eta^{\circ}$ | $\dfrac{\tilde{\eta}_1}{\tilde{\eta}_2} = \dfrac{\tilde{\eta}_1}{\tilde{\eta}_{s2} + \tilde{k}\left(\sqrt{\frac{\tilde{g}}{\tilde{R}}}\right)^{n-1}}$ | Viscosity of the drop and total viscosity of the EVP material |
| Archimedes, $Ar$ | $\dfrac{(\rho^{\circ}-1)\tilde{\rho}_2\sqrt{\tilde{g}\tilde{R}}\tilde{R}}{\tilde{\eta}_2}$ | Buoyancy and viscous stresses |
| Bond, $Bo$ | $\dfrac{(\rho^{\circ}-1)\tilde{\rho}_2\tilde{g}\tilde{R}^2}{\tilde{\Gamma}}$ | Buoyancy and capillary stresses |
| Bingham, $Bn$ | $\dfrac{\tilde{\sigma}_y}{(\rho^{\circ}-1)\tilde{\rho}_2\tilde{g}\tilde{R}}$ | Plastic (yield) and buoyancy stresses |
| Elastogravity, $Eg$ | $\dfrac{(\rho^{\circ}-1)\tilde{\rho}_2\tilde{g}\tilde{R}}{\tilde{G}}$ | Buoyancy and elastic stresses |
| Viscosity ratio in EVP, $\beta$ | $\dfrac{\tilde{\eta}_{s2}}{\tilde{\eta}_2} = \dfrac{\tilde{\eta}_{s2}}{\tilde{\eta}_{s2} + \tilde{k}\left(\sqrt{\frac{\tilde{g}}{\tilde{R}}}\right)^{n-1}}$ | Solvent viscosity to total viscosity in the EVP material |

The corresponding dimensionless equations are:

$$\rho(\phi) = \frac{\tilde{\rho}}{\tilde{\rho}_2} = \phi + \rho^{\circ}(1-\phi) \tag{17}$$

$$\eta_s(\phi) = \frac{\tilde{\eta}_s}{\tilde{\eta}_2} = \beta\phi + \eta^{\circ}(1-\phi) \tag{18}$$

$$\eta_p(\phi) = \frac{\tilde{\eta}_p}{\tilde{\eta}_2} = (1-\beta)\phi \tag{19}$$

$$\boldsymbol{u} = \boldsymbol{u_2}\,\phi + \boldsymbol{u_1}\,(1-\phi) \tag{20}$$

$$\boldsymbol{\tau_p} = \boldsymbol{\tau_{p,2}}\,\phi + \boldsymbol{\tau_{p,1}}\,(1-\phi) \tag{21}$$

$$\nabla\cdot\boldsymbol{u} = 0 \tag{22}$$

$$\left(\frac{\partial\boldsymbol{u}}{\partial t} + \boldsymbol{u}\cdot\boldsymbol{\nabla}\boldsymbol{u}\right) = \frac{\rho^{\circ}-1}{\rho(\phi)}\left[-\boldsymbol{\nabla}P + \frac{1}{Bo}\kappa\boldsymbol{n}\delta + \frac{\eta_s(\phi)}{Ar}\boldsymbol{\nabla}^2\boldsymbol{u} + \boldsymbol{\nabla}\cdot\boldsymbol{\tau_p}\right] - \boldsymbol{e_z} \tag{23}$$

$$Eg\,\overset{\nabla}{\boldsymbol{\tau}}_{\boldsymbol{p},2} + \max\left(0, \frac{Ar}{1-\beta}\left(\left|\boldsymbol{\tau_{p,d,2}}\right| - Bn\right)\right)^{\frac{1}{n}}\frac{\boldsymbol{\tau_{p,2}}}{\left|\boldsymbol{\tau_{p,d,2}}\right|} = \left(\boldsymbol{\nabla}\boldsymbol{u_2} + (\boldsymbol{\nabla}\boldsymbol{u_2})^{\boldsymbol{T}}\right) \tag{24}$$

$$\frac{\partial\phi}{\partial t} + \boldsymbol{\nabla}\cdot(\phi\boldsymbol{u}) = 0 \tag{25}$$

### *2.3 Fluid rheology*

We opt for a moderately concentrated Carbopol solution in water to represent the YSM in our study. This choice stems from Carbopol's widespread use in experimental settings when a yield stress material is required. Notably, an increase in polymer concentration accentuates elastic, plastic, and thixotropic contributions. Our aim is to replicate the experimental observations of Lavrenteva, Holenberg, and Nir (hereafter referred to as LHN) [35]. LHN conducted experiments on the sedimentation of tetrachloroethylene drops in a cylindrical tube filled with neutralized Carbopol at varying concentrations (0.07%, 0.08%, 0.09%). For our investigation, we primarily focus on the solution at 0.07% concentration. LHN's subsequent studies using the same material report the emergence of thixotropic effects at higher concentrations [34], [36]. However, our study exclusively centers on analyzing the interplay between plastic and elastic effects. LHN characterized Carbopol's rheology through steady shear experiments, concluding that the Herschel-Bulkley constitutive equation adequately models the material, considering elastic effects negligible. Nonetheless, our examination of drop shapes and analysis of both flow fields and yield surfaces indicate the necessity of incorporating elastic effects to provide a comprehensive description and understanding of the observations. Since the authors did not include either the steady shear curve or the strain amplitude / frequency sweep curves, we have selected two reasonable values of the elastic modulus $\tilde{G}$ and the shear-thinning parameter $n$, based on other experimental works involving Carbopol at similar concentrations [23], [24], while the consistency index $\tilde{k}$ and the yield stress $\tilde{\sigma}_y$ are fixed to the same values employed in LHN. The solvent viscosity in the elastoviscoplastic phase, $\tilde{\eta}_{s_2}$, is chosen to obtain in all cases a ratio with the total EVP viscosity of $\beta = 0.1$. Such small, but non null, value is chosen to enhance the numerical stability, by preserving the ellipticity of the momentum balance, yet ensuring that the solvent contribution is small with respect to its polymeric counterpart.

We consider drops with an effective radius around 3 mm, and the terminal velocities observed in experiments are $O\left(10^{-2}\,\frac{\text{m}}{\text{s}}\right)$, hence we expect the characteristic shear rate to be $\tilde{\dot{\gamma}}_{ch} \approx 3\ \left[\frac{1}{\text{s}}\right]$. As depicted in **Fig. 3**, our steady predictions align quantitatively with the experimental data in this region. However, it is widely acknowledged that under complex flow conditions, steady shear data alone are often insufficient to properly characterize the rheological response of the material [40]; this uncertainty is further compounded by the lack of experimental data regarding the extensional properties of the material and its transient response to start-up flow, which requires careful consideration.

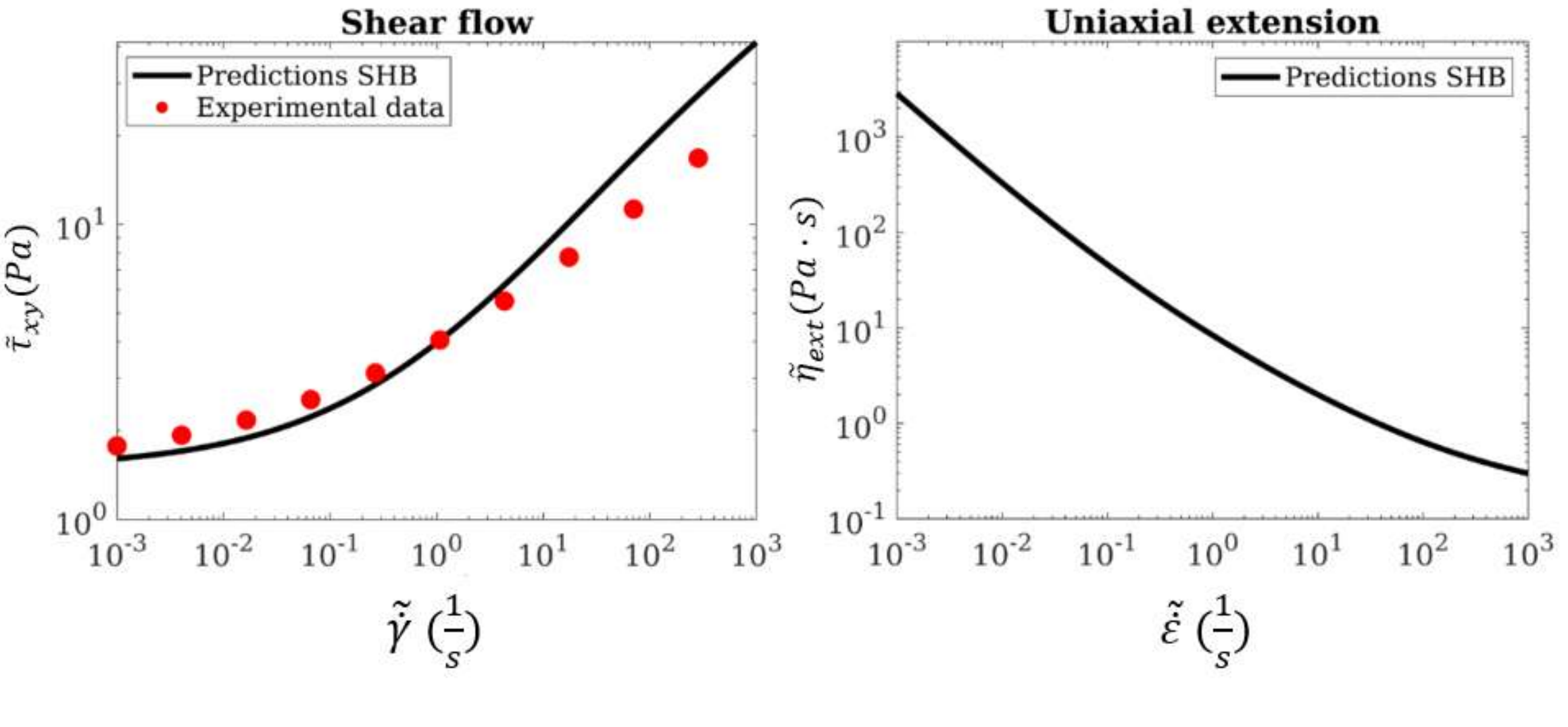

**Fig. 3:** Steady shear data (left) and uniaxial extension (right). The experimental data correspond to the predictions of the Herschel-Bulkley model according to LHN [35]

Having this in mind, we explore the effect of the rheological parameters in § 3.4, to justify possible discrepancies with the experimental reference. The properties of the drop are the ones of tetrachloroethylene at 20°C, and the interfacial tension $\tilde{\Gamma}$ is selected to be equal to the one of the system tetrachloroethylene-water at the same temperature [35]. This assumption is supported by the low concentration of Carbopol employed for the base case, since several studies in the literature confirm that water and Carbopol have similar values of $\tilde{\Gamma}$ when the concentration of the latter is relatively low [41]. All the material properties for both the Newtonian and the EVP phases are listed in **Table II**.

**Table II:** Rheological parameters for the YSM and the Newtonian drop in the base case.

| ***Symbol*** | ***Value*** |
|---|---|
| $\tilde{\sigma}_y$ | 1.5 [Pa] |
| $\tilde{k}$ | 2.5 [Pa · s$^n$] |
| $\tilde{G}$ | 32 [Pa] |
| $n$ | 0.45 |
| $\tilde{\rho}_2$ | 1000 [kg/m$^3$] |
| $\tilde{\rho}_1$ | 1621 [kg/m$^3$] |
| $\tilde{\eta}_1$ | $8.9 \cdot 10^{-4}$[Pa · s] |
| $\tilde{\Gamma}$ | 0.044 [N/m] |

### *2.4 Numerical method*

We perform the numerical simulations employing the open-source finite volume solver Basilisk [42]. This solver has been extensively validated for free surface flows of Newtonian fluids [43], [44], and recently extended to simulate flow problems involving viscoelastic materials [45], [46], introducing well-known stabilization techniques that allow to simulate flow problems at high $Wi$ number, i.e., when elastic effects are dominant [47]. The system of partial differential equations is solved on a Cartesian grid with a collocated discretization of the velocity, pressure, and stress fields. An adaptive mesh refinement technique (AMR) based on a wavelet decomposition is employed to locally refine the grid [48]. We choose to refine the computational cells according to the estimated error for the velocity, the stress, and the indicator function $\phi$, since the regions where we desire the maximum accuracy are the interface of the drop and the yield surface. Such procedure requires the specification of three parameters regulating the refinement level of the grid, namely the initial level $N$, the maximum level $N_{MAX}$ and the minimum level $N_{MIN}$. Such parameters identify the characteristic grid size $\Delta\tilde{x}$ according to the relation $\Delta\tilde{x} = \frac{\tilde{L}}{2^N}$. To minimize the error associated with the definition of the volume of the drop, we initially refine the whole domain $\Omega_1$. The initial finer mesh inside and near the drop can be seen in **Fig. 4**(a). At later times with deformed drops and complicated yield surfaces around them, the mesh is properly refined as described above and seen in **Fig. 4** (b). The time integration is based on a second-order fractional step method [49], where the maximum time step is selected according to the capillary timescale in order to reduce the occurrence of fake capillary waves [50], imposing $\max(\widetilde{\Delta t}) \leq \sqrt{\frac{\tilde{\rho}_{avg}\widetilde{\Delta x}^3}{\pi\tilde{\Gamma}}}$, with $\tilde{\rho}_{avg} = \frac{\tilde{\rho}_1+\tilde{\rho}_2}{2}$. To ensure that our calculations are independent on the grid size and the selected time step, we perform a mesh and time convergence study reported in the Appendix. Consequently, we select N=10, $N_{MAX} =$ $12, N_{MIN} = 6$. This implies that for a value of $L = 40$ and $N = 10$, the initial (uniform) grid includes square-shaped cells whose side has a length of $\Delta x = \frac{40}{2^{10}} \approx 0.04$, while the value $N_{max} = 12$ corresponds to a cell-side, in the region of maximum refinement, of approximately 0.01 (i.e., 100 cells per initial radius). The value of $N_{min}$ is related to the size of the cells in the unrefined region, far away from the interface of the drop and the yield surface. Please notice that all the figures are mirrored with respect to the axis of symmetry to report the complete shape of the drop.

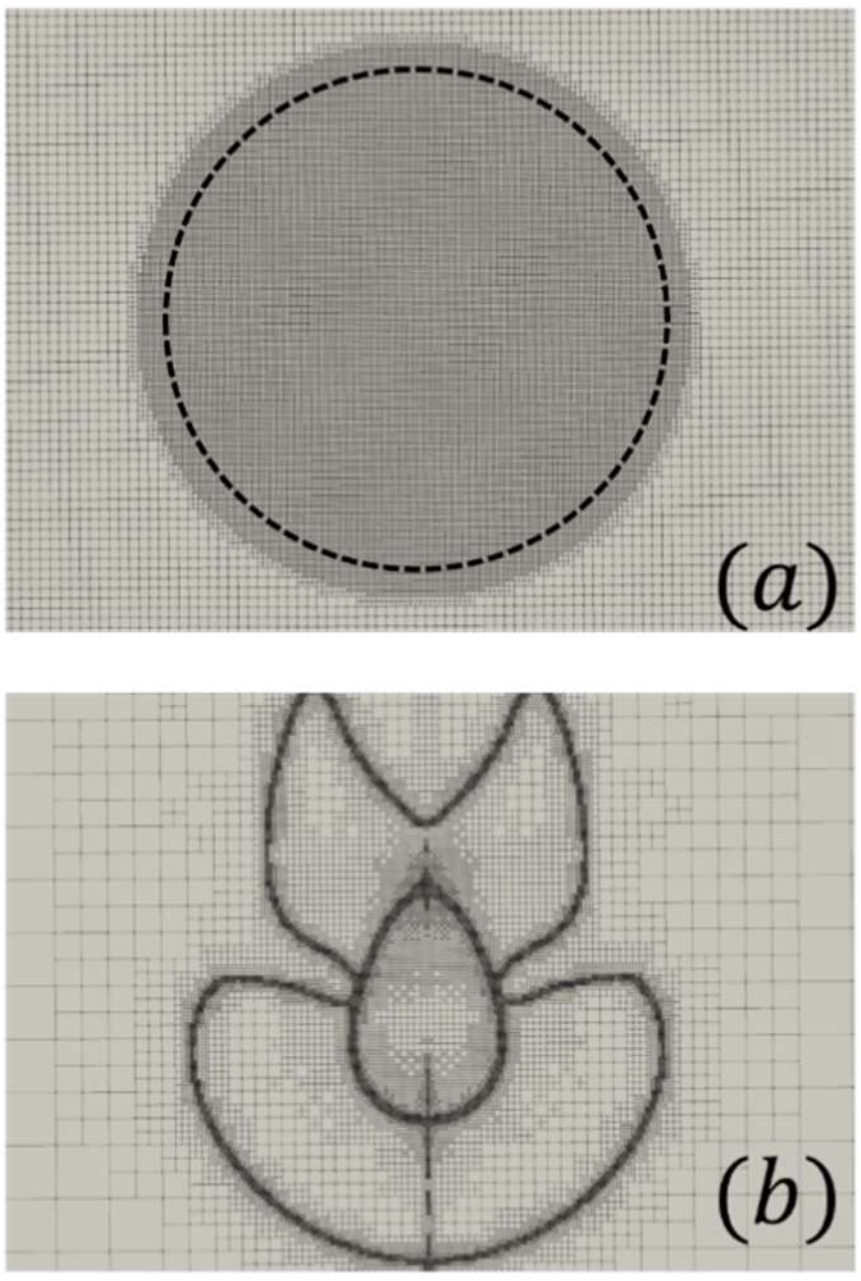

**Fig. 4:** Mesh resolution in the proximity of the drop interface and the yield surface at the start of the simulation (a), and after a steady shape is attained (b). The interface separating the initially spherical drop and the YSM in (a) is given by a dotted line. Please notice that along the axis of symmetry we adopt the maximum level of refinement to accurately capture the evolution of the normal stresses.

### 3. Results and discussion

In this section, we present and discuss the results obtained from our study. Firstly, we conduct validation tests to ensure the model appropriateness and the adequacy of the numerical solver utilized in this investigation. Subsequently, we compare our findings with the experimental results of LHN, focusing on the shape and velocity of the drops. Lastly, we delve into an extensive parametric study to explore the effect of rheological parameters on sedimentation dynamics. To facilitate comprehension, we provide definitions for several relevant quantities essential for result interpretation: the axial position of the center of mass, denoted as $z_c$, is computed as the volume integral of all axial positions, weighted by the complementary indicator function $(1 - \phi)$. Similarly, velocity is determined through volumetric integration of the local velocity field, employing the complementary indicator function as the weight. Furthermore, we define the Taylor parameter $D$ to quantify the deformation of the drop during its sedimentation.

$$z_c = \frac{\int_V z(1-\phi)dV}{\int_V (1-\phi)dV}, \quad u_c = \frac{\int_V u_z(1-\phi)dV}{\int_V (1-\phi)dV} \tag{26}$$

$$D = \frac{H-W}{H+W} \tag{27}$$

Here, H (height) and W (width) are the values of the maximum extension of the drop, respectively, in the axial and the radial direction. Consequently, for a spherical object $D = 0$, for a prolate shape $D > 0$ and for an oblate shape this value is negative. Finally, we represent the interface as the locus of the points corresponding to $\phi = 0.5$.

### *3.1 Code validation.*

In order to validate the model and its numerical implementation, we make an effort to reproduce the numerical results of Moschopoulos et al. [30], who studied numerically the rising of an air bubble in a YSM modelled via the Saramito-Herschel-Bulkley constitutive equation by means of a finite element formulation, tracking the interface via the Arbitrary Lagrangian-Eulerian (ALE) method; a completely different methodology. The authors compare their numerical predictions with two experimental references [23], [24], finding quantitative agreement in terms of bubble shapes and terminal velocities. The YSM material corresponds to a solution of Carbopol (concentration 0.10%) in water, whose rheological parameters are extracted via non-linear fitting of the flow curves reported in the experimental references and reported in Table III.

**Table III:** Rheological parameters for the modelling of the YSM material via the Saramito-Herschel-Bulkley constitutive equation in the validation tests for a rising bubble.

| ***Symbol*** | ***Value*** |
|---|---|
| $\tilde{\sigma}_y$ | 4.71 [Pa] |
| $\tilde{k}$ | 1.81 [$\mathrm{Pa \cdot s^n}$] |
| $n$ | 0.46 |
| $\tilde{G}$ | 40.42 [Pa] |
| $\tilde{\Gamma}$ | 0.073 [N/m] |
| $\tilde{\rho}_2$ | 1000 [$\mathrm{kg/m^3}$] |

The density and viscosity of the bubble in [30] are considered to be negligible with respect to the corresponding values in the elastoviscoplastic fluid. In our case, we select the representative values of $\tilde{\rho}_1 = 1\ \mathrm{kg/m^3}$ and $\tilde{\eta}_1 = 1.2 \cdot 10^{-5}\ \mathrm{Pa \cdot s}$.

We commence the validation process by comparing the terminal shape of a bubble of $\tilde{R} = 4$ mm, along with the extent of the yielded region, in **Fig. 5**. As evident from the figure, our numerical setup accurately reproduces the inverted teardrop shape acquired in [30]. A slight discrepancy in the shape of the yield surface is attributed to the fact that in [30] the value of the solvent viscosity in the elastoviscoplastic phase is null. Furthermore, the yield surface predicted in this study is smoother due to the adaptive mesh refinement around it, whereas in [30] the mesh was monotonically coarsened.

We proceed replicating the prediction for a bubble of $\tilde{R} = 8.3$ mm, where an oblate shape is recovered, a sign of prevailing inertia over elasticity, in **Fig. 6**. Again, we find a satisfactory agreement concerning the terminal shape. Again, the yield surface in our study is smoother than the one shown in the numerical reference, due to the fact that we refine the cells in proximity of the transition region.

The shape of the interface for two other cases, with, respectively, $\tilde{R}$ = 10.7 mm and $\tilde{R}$ = 16.3 mm, is compared in **Fig. 7**. Here, an oblate shape is recovered due to the dominating inertial effects. In this case also, the agreement in terms of bubble shapes is satisfactory.

We also report the transient velocity profiles and compare their final steady values with the terminal velocities in all four cases in **Fig. 8**. The results are summarized in **Table IV**. The error in the prediction of the terminal velocity is below 2 % in all cases.

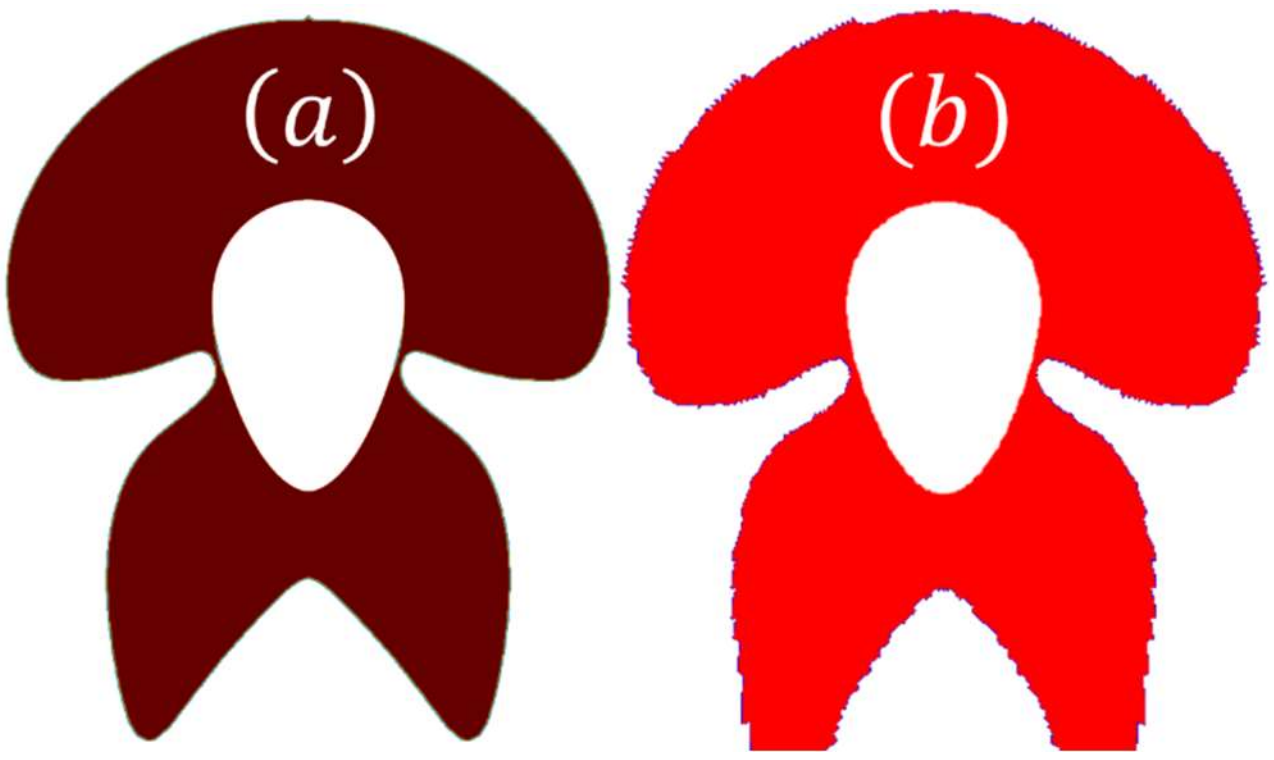

**Fig. 5:** Test case for a bubble of $\tilde{R}$ = 4 mm. Comparison between the terminal shape predicted in the present study (a) and the corresponding shape predicted in [30] (b). The bordeaux region in (a) and the red region in (b) represent the yielded region, where the YSM behaves as a shear thinning viscoelastic liquid. Please notice that the yield surface predicted in our study is smoother than the one shown in the numerical reference, due to the fact that we refine the cells in the proximity of the transition region.

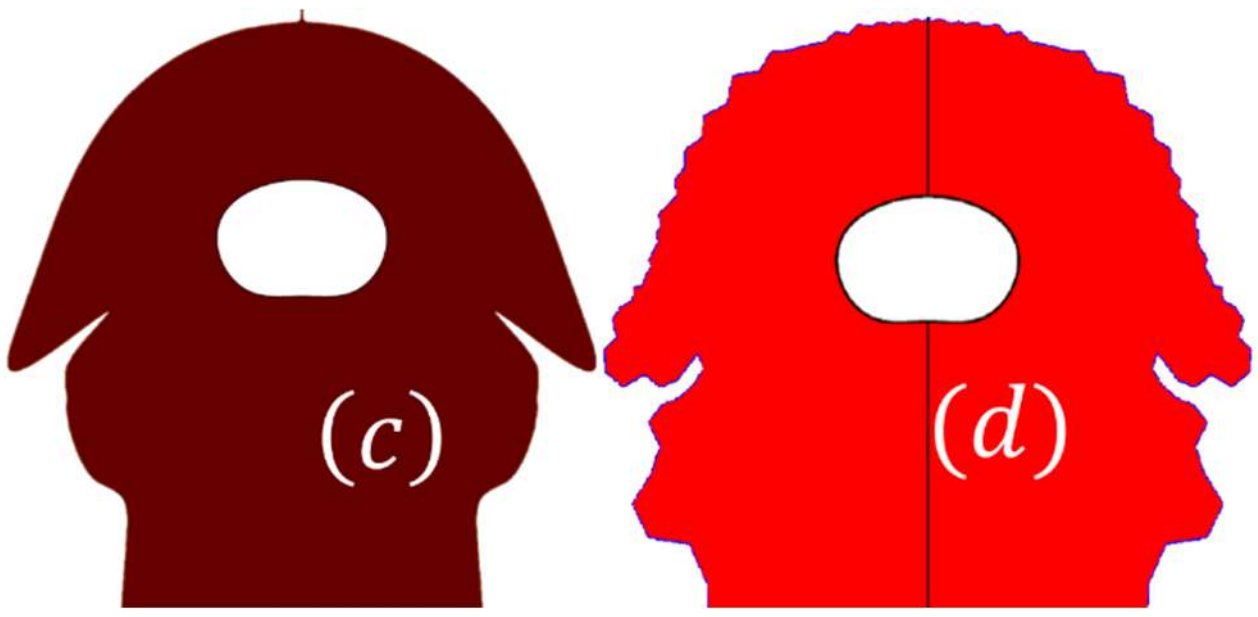

**Fig. 6:** Test case for a bubble of $\tilde{R}$ = 8.3 mm. Comparison between the terminal shape predicted in the present study (c) and the corresponding shape predicted in [30] (d). The bordeaux region in (c) and the red region in (d) represent the yielded region.

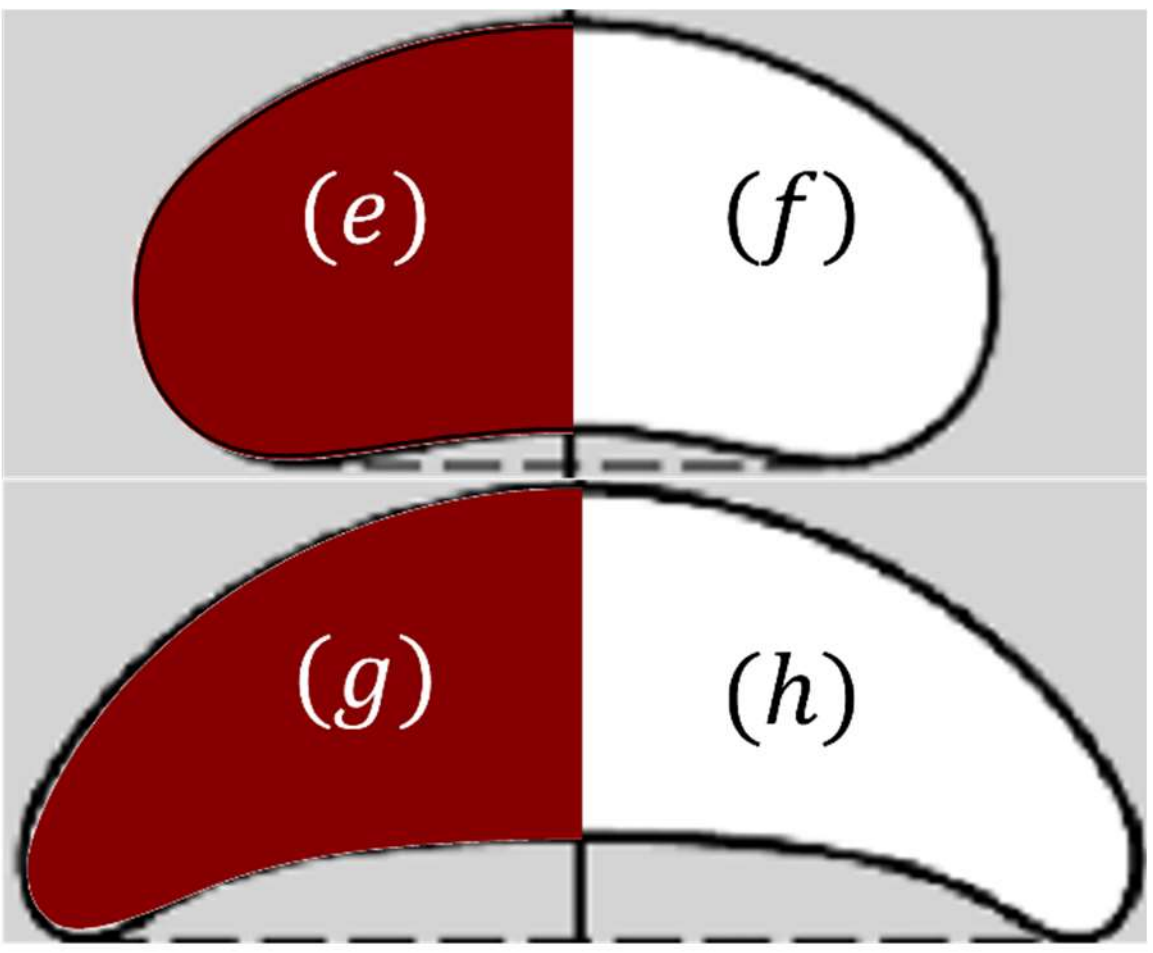

**Fig. 7:** Comparison of the terminal shape for bubbles of $\tilde{R}$ = 10.7 mm and $\tilde{R}$ = 16.3 mm. Our results are respectively (e) for a bubble of $\tilde{R}$ = 10.7 mm and (g) for a bubble of $\tilde{R}$ = 16.3 mm. The corresponding predictions in [30] are indicated with (f) and (h).

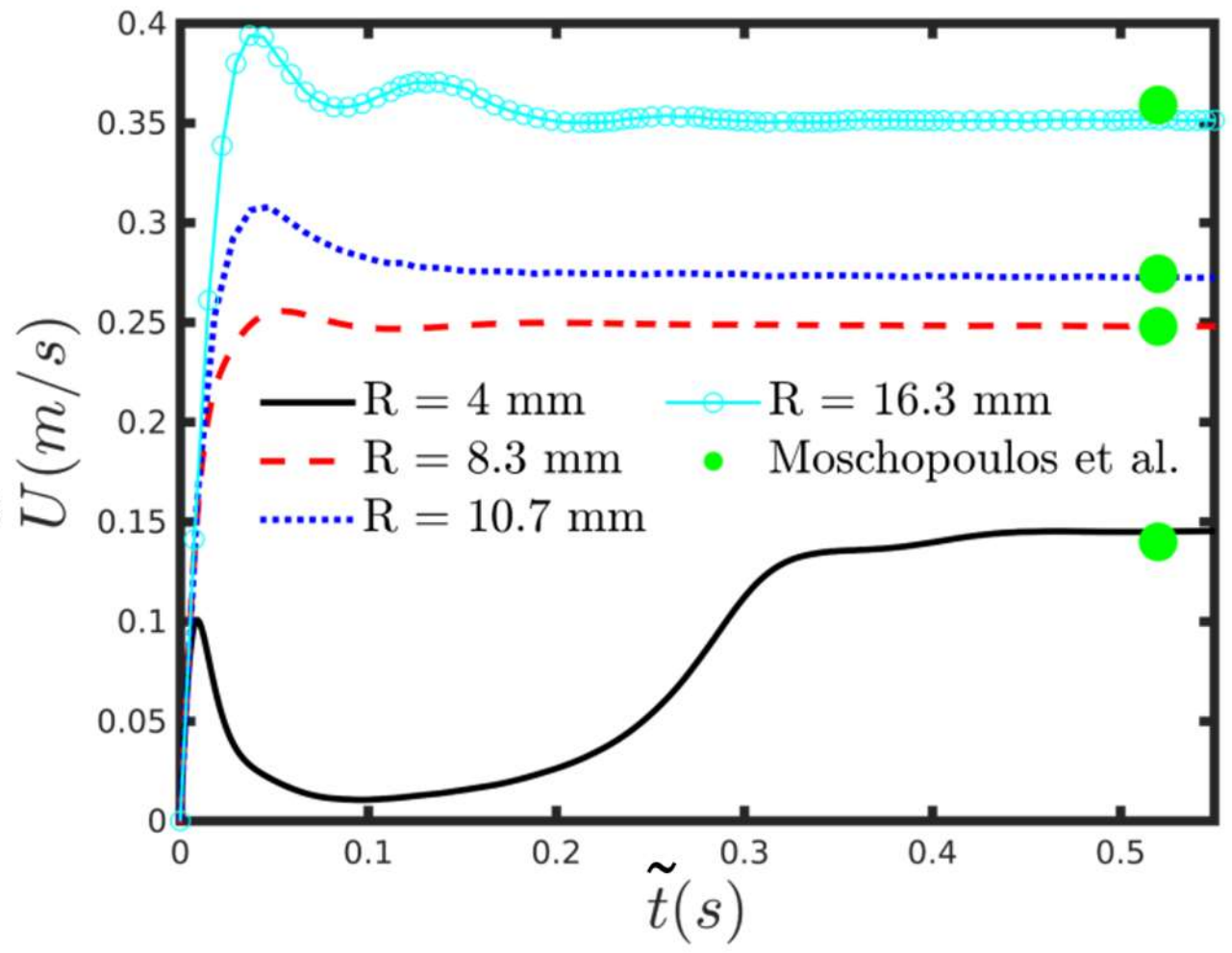

**Fig. 8:** Transient velocity profiles for four values of the effective radius and comparison with the steady velocities reported in [30].

**Table IV:** Comparison of the predicted terminal velocities for bubbles of four different radii with the ones reported in [30].

| **Radius (mm)** | $\tilde{U}_t$ **(m/s) in** [30] | $\tilde{U}_t$ **(m/s) in this study** | **Error (%)** |
|---|---|---|---|
| 4 | 0.145 | 0.144 | 0.68 |
| 8.3 | 0.248 | 0.247 | 0.40 |
| 10.7 | 0.2744 | 0.2706 | 1.38 |
| 16.3 | 0.359 | 0.354 | 1.39 |

### *3.2 Comparison with experiments*

Besides the model validation, we are also interested in the comparison between our numerical predictions and the experimental results reported by [35]. We compare the values of the terminal velocity and shape obtained by the sedimenting drop. Since all the dimensionless numbers are affected by the effective radius of the drop, it is instructive to characterize every case corresponding to a different effective radius with its own set of dimensionless numbers. For the following results, the density ratio is fixed to 1.621, while the viscosity ratio depends on the effective characteristic viscosity in the YSM, which depends on the radius. For a drop of effective radius $\tilde{R}$=3.6 mm (equivalent volume of approximately 200 μL), the corresponding dimensionless numbers are $Ar = 1.46,\ Bo = 1.79,\ Bn = 0.068,\ Eg = 0.68,\ \eta^{\circ} = 0.003, \beta = 0.1$. Our predicted shape matches quite well with the experimental one, as reported in **Fig. 9**.

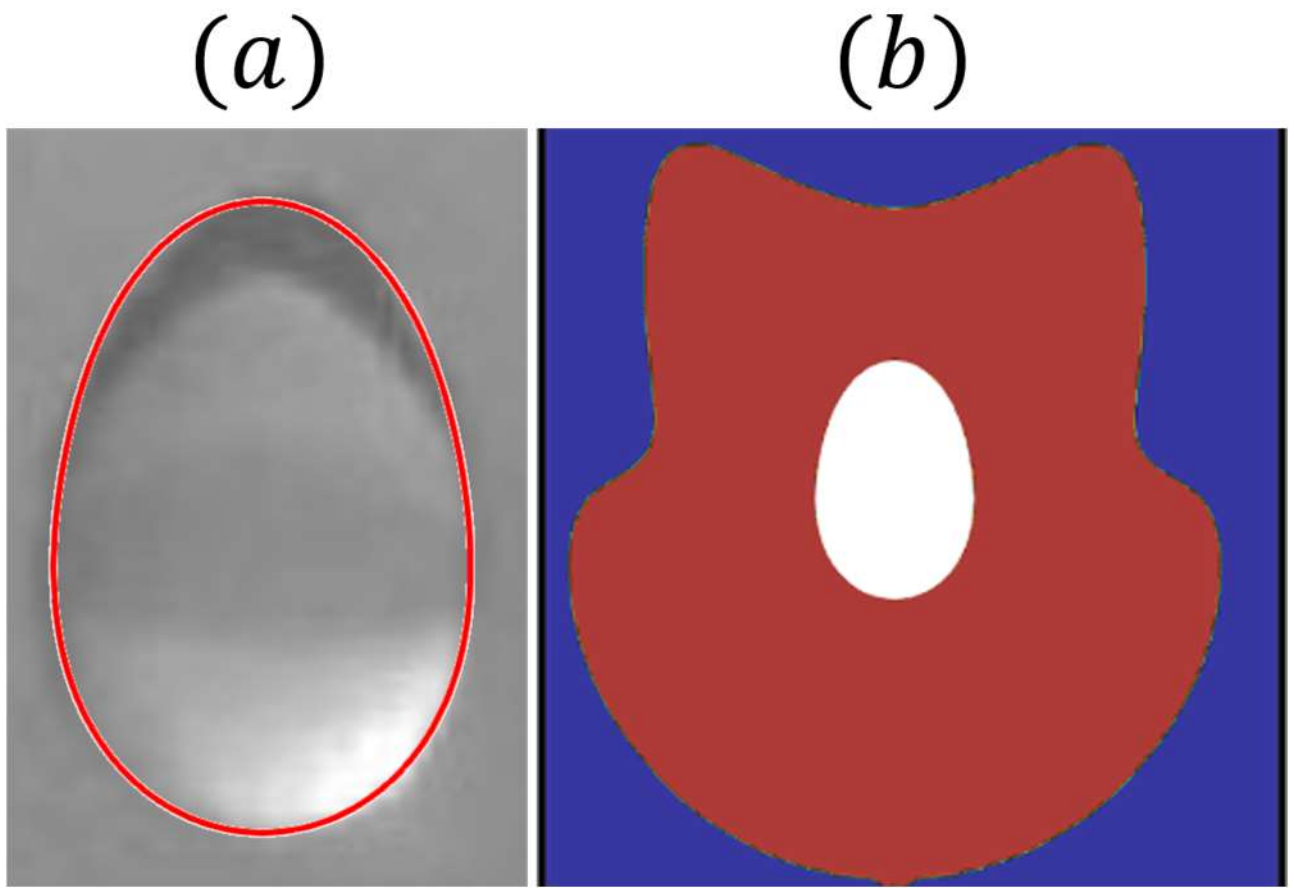

**Fig. 9:** Drop of effective radius $\tilde{R} = 3.6$ mm, (a) comparison between experimentally reported and numerically predicted terminal shape (red line). (b) the yielded region is indicated in red, the drop in white and the unyielded region in blue.

The velocity evolution is reported in **Fig. 10**. We compute the corresponding Reynolds number as $Re = Ar\ U_t$, with $U_t$ the dimensionless terminal sedimentation velocity of the drop, to evaluate the relevance of inertial effects. In this case, the terminal sedimentation velocity is equal to $U_t = 0.056$, with a corresponding $Re = 0.084$. Such low value of $Re$ indicates that the slightly elongated shape acquired by the sedimenting drop cannot be caused by inertial effects.

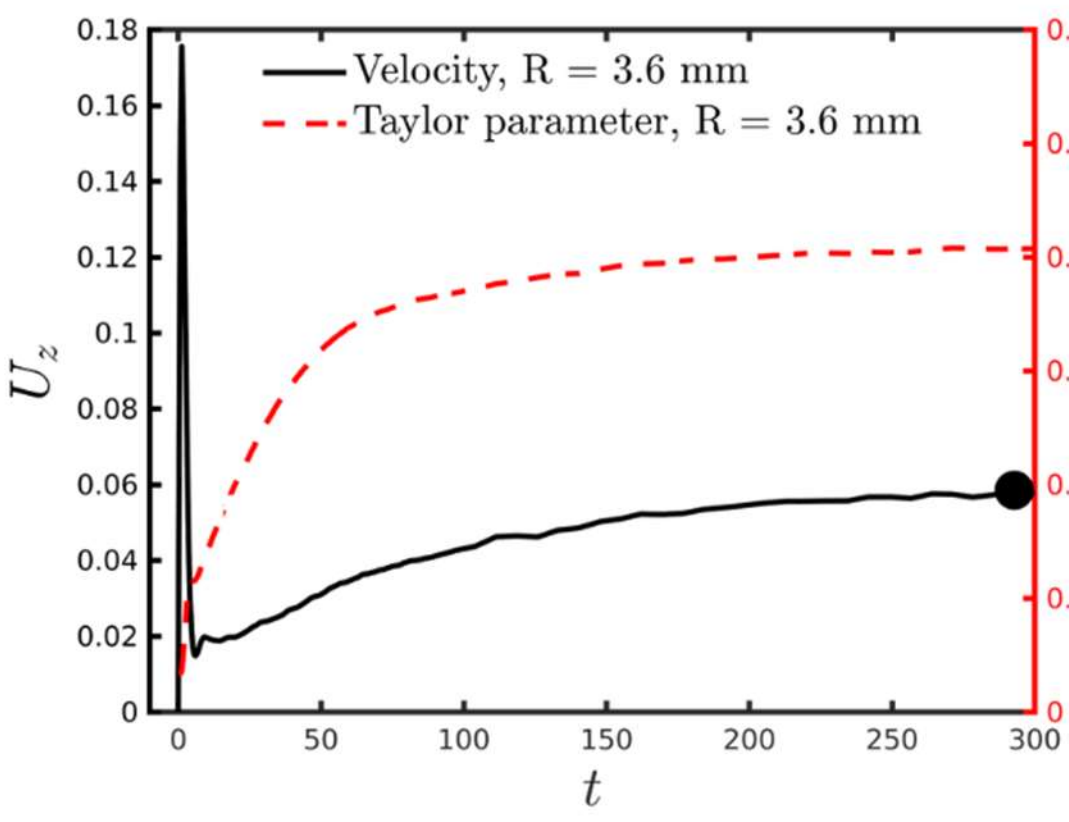

**Fig. 10:** Time evolution of the sedimentation velocity and Taylor deformation parameter $D$ for a drop with effective radius $\tilde{R} = 3.6$ mm. The black circle indicates the terminal velocity

reported in the experiments of LHN. The positive value of the sedimentation velocity $u_z$ indicates movement of the drop toward the bottom container (i.e., in the negative $z$ direction of the coordinate system).

Previous numerical simulations concerning the buoyancy driven motion of bubbles in viscoplastic materials [21] showed that the aspect ratio of the deformable intrusion for a similar set of dimensionless numbers is well below unity. In such cases, the Taylor parameter is negative and the bubble acquires an oblate shape with no rounded tip at the rear. This deviation confirms that elasticity is crucial to generate the elongated shapes correctly.

The rounded tip at the rear of the sedimenting drop is generated by the coexistence of high extensional and shear stresses, as displayed in **Fig. 11**. Furthermore, the analysis of the axial velocity distribution in the proximity of the interface highlights another typical elastic effect: the insurgence of a negative wake, i.e. an inversion of the flow-direction behind the sedimenting object, which has also been experimentally reported via PIV measurements [36]. Such phenomenon has been previously reported for the buoyancy driven motion of spheres [51], drops [52] and bubbles [29] in viscoelastic solutions. The shape of the yielded region is in qualitative agreement with the one observed around a solid sphere with a smooth surface when elastic effects are included [16].

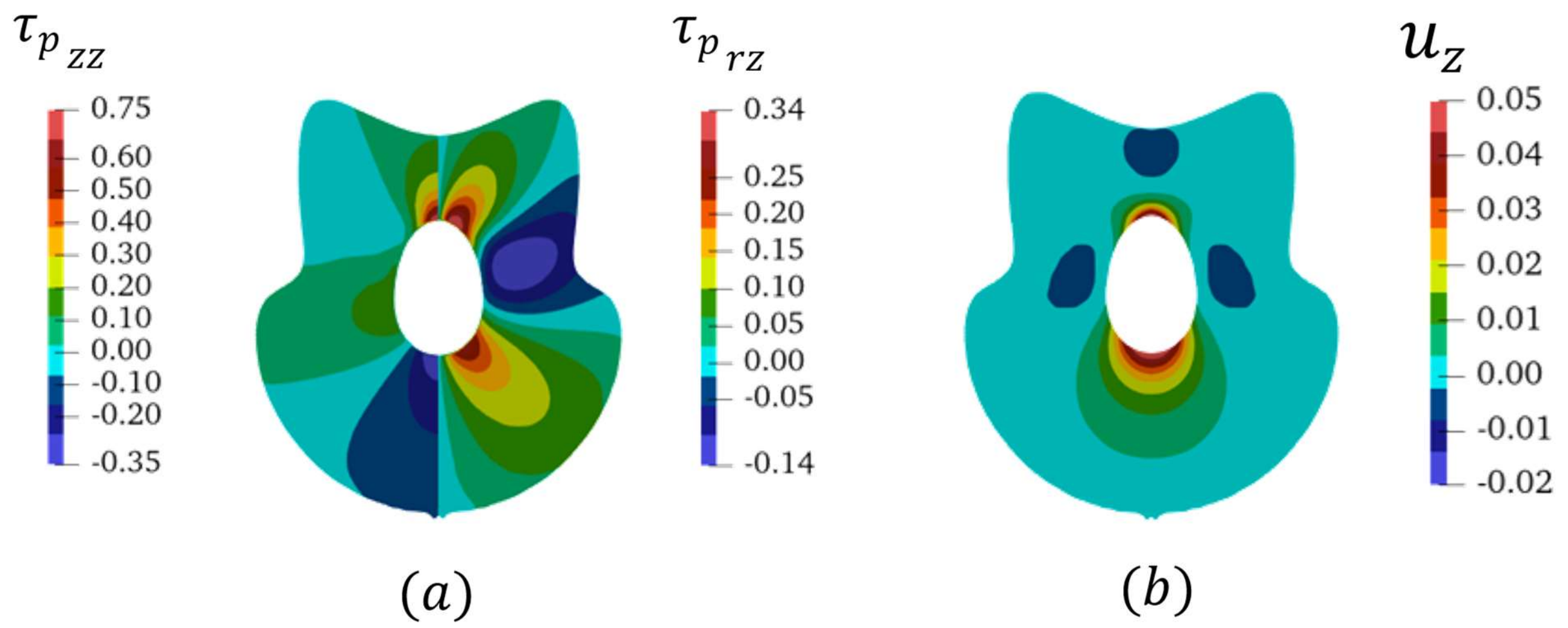

**Fig. 11:** For a drop of effective radius $\tilde{R}$ = 3.6 mm, the contours of $\tau_{p_{zz}}$ and $\tau_{p_{rz}}$ are reported respectively in (a, left) and (a, right). The axial sedimentation velocity is displayed in (b). Notice that positive values of $u_z$ correspond to a movement of the drop toward the negative $z$ direction of the coordinate system.

On the contrary, a deviation from previous numerical simulations involving drops and bubbles moving in YSM modelled through inelastic constitutive equations (i.e., Bingam, or Herschel-Bulkley) arises [37], [53]. Potapov et al. [37] studied the sedimentation of a viscous drop in an inelastic YSM modelled through the regularized Bingham law. It is instructive to compare our results, when elasticity is included in the description of the continuous phase, with their predictions, see **Fig. 12**. In both cases, inertial effects are subdominant. In **Fig. 12** (a), the deformability of the sedimenting drop allows the interface to take a slightly prolate, non symmetric shape with an indentation at its rear, but the resulting yielded domain depicted with green-yellow is fore-aft symmetric, as expected in the buoyancy driven motion of objects in an inelastic fluid under creeping flow conditions. On the other hand, for a similar $Bn$, our numerical predictions show the insurgence of a fore-aft asymmetry in the velocity and stress distributions, as observed in **Fig. 11**, resulting in an asymmetry in the drop shape and yield surface in **Fig. 12** (b).

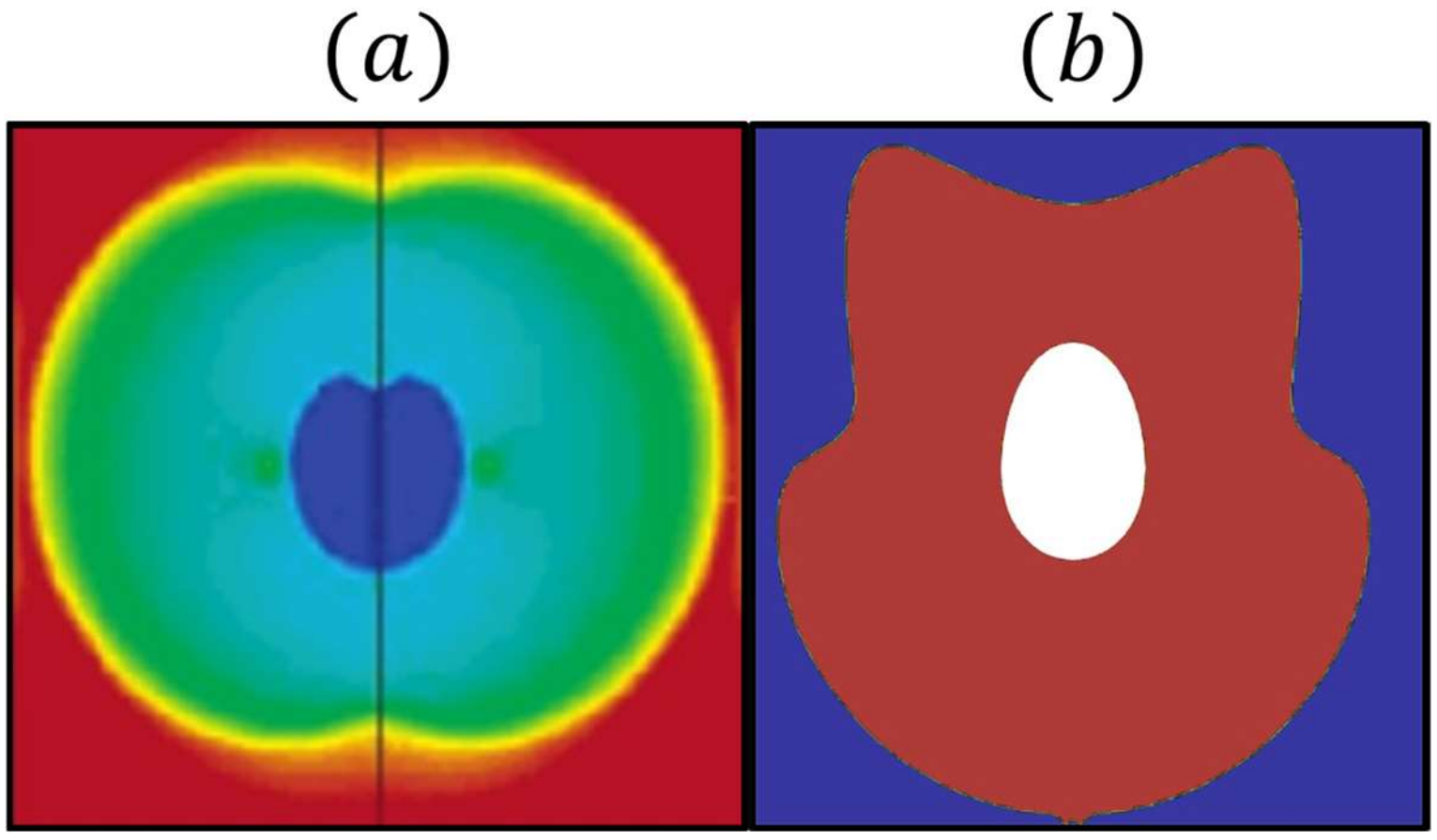

**Fig. 12:** Comparison between (a) the numerical predictions of Potapov et al. [37] for a drop sedimenting in a Bingham fluid at $Bn = 0.058$ and (b) our numerical predictions employing the SHB constitutive equation, at $Bn = 0.068$. The color scale in (a) reflects the value of the plastic viscosity, where the yielded region assumes a peach-like shape and the unyielded region is indicated in red. In our predictions, the yielded region is indicated in red, while the unyielded region is shown in blue.

A thorough characterization of flow conditions around a moving object in YSMs necessitates precise evaluation of the yield surface, delineating the boundary between solid-like and fluid-like behavior in the continuous phase. The von Mises criterion is a generally accepted and reliable method for identifying this separatrix, albeit it requires assessing the stress distribution to compute the second invariant of the deviatoric stress tensor. Previous studies have attempted to "track" the yield surface by monitoring the local fluid velocity [14], [36]. Indeed, for inelastic YSMs, the deviatoric stress tensor can be directly correlated with the rate of deformation tensor. Measurement of the rate of deformation tensor can be achieved through standard techniques such as Particle Image Velocimetry (PIV) or Particle Tracking Velocimetry (PTV) [14], [36]. With this approach, the material is assumed to be fluidized (i.e. *yielded*) when the magnitude of the velocity is higher than a threshold, usually around 5% of the drop velocity [36]. In **Fig. 13**, we report the contours of the velocity magnitude and the yield surface calculated according to the von Mises criterion, with the aim to compare the distribution of *yielded* and *unyielded* regions obtained through a stress-based criterion with the ones obtained through the measurement of the velocity field. Having in mind that we represent the isocontours of the velocity magnitude in the range $\frac{|\tilde{U}|}{\tilde{U}_t} \in [0.001 - 0.006]$, we observe a clear mismatch between the isocontours of the velocity magnitude and the yield surface calculated through the von Mises criterion (white contour in **Fig. 13**). Such mismatch has been observed also in the case of solid spheres sedimenting in YSMs [16] and clearly indicates that the measurement of the velocity field is necessary, but not sufficient to accurately detect the transition region. Indeed, the calculation of the stress distribution is crucial for the computation of the second invariant of the deviatoric stress tensor and for the application of the von Mises criterion. Furthermore, the streamlines display the occurrence of the negative wake right behind the rear of the sedimenting drop. We also notice that the flow is directed upward as well for a small region outside the yield surface and around the equatorial plane of the drop, due to the elastic response of the material in the unyielded region.

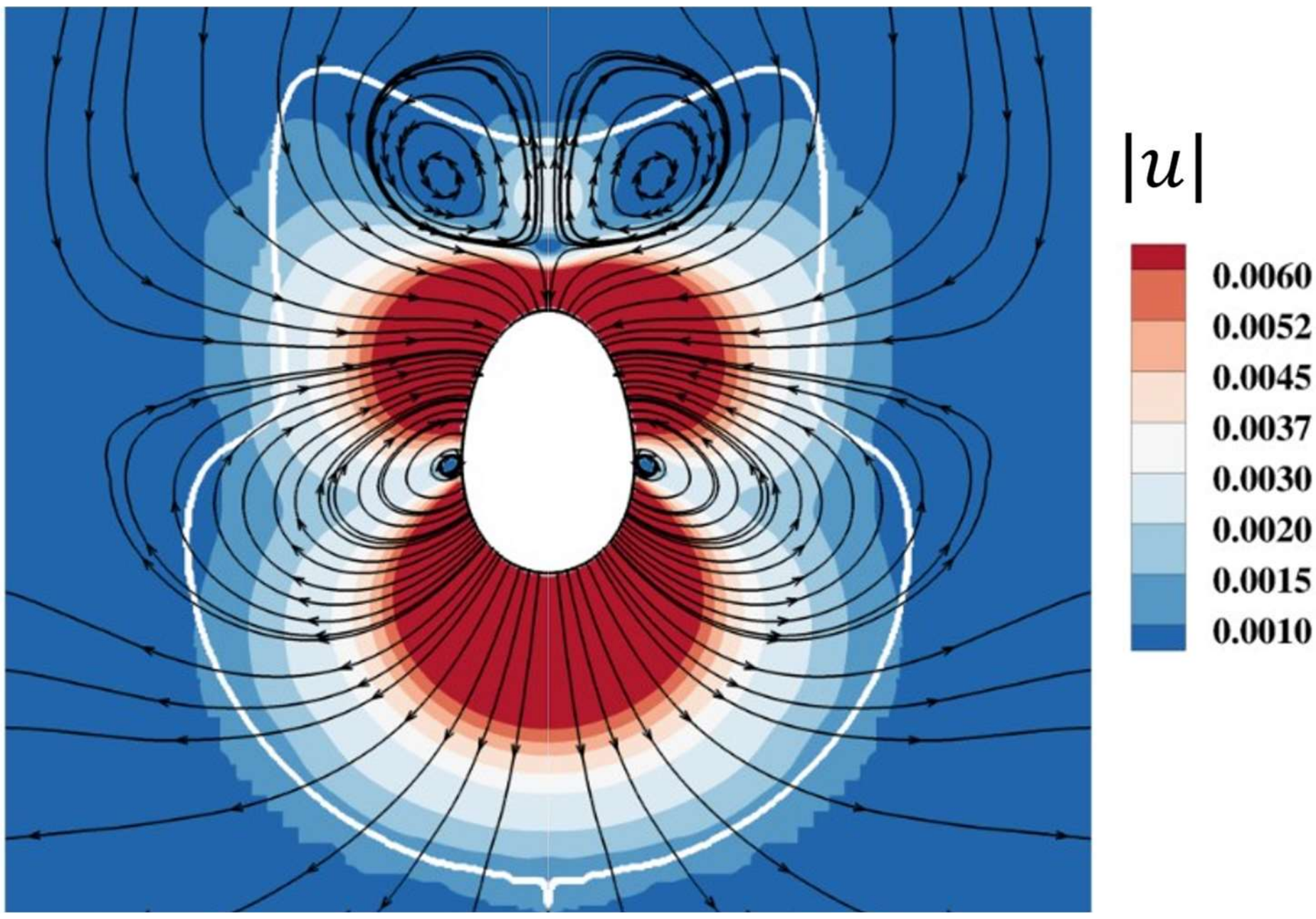

**Fig. 13:** Contours of the velocity magnitude around the sedimenting drop and yield surface (white line) calculated according to the von Mises criterion.

In **Fig. 14**, we present the time development of the sedimentation velocities (taken in their absolute values) and the deformation parameters for drops of different radii. For the cases where the steady sedimentation velocity has been reported experimentally, we compare these values with the terminal velocities predicted by our numerical simulations and summarize the comparison in **Table V**. For all the radii, the initial stage of motion is characterized by the interplay between the buoyancy force, the viscous resistance and the plastic response of the material. The material is initially devoid of viscoelastic stresses ($\boldsymbol{\tau_p}|_{t=0} = \mathbf{0}$). However, the initial motion of the drop triggers the accumulation of these stresses, leading to a strong overshoot at the very early stage of motion. Subsequently, various scenarios unfold depending on the effective radius: smaller drops gradually approach a steady velocity, whereas larger drops experience a secondary acceleration before reaching a stable terminal velocity, which may even surpass the initial overshoot. Such behaviour has been previously reported also for bubbles rising in EVP materials [30] or for initially static drops that start rising in viscoelastic solutions [52]. Examining the evolution of the Taylor deformation parameter $D$ in **Fig. 14** (b), we deduce that the second acceleration of the drops corresponds to the development of the elongated shape, the latter is caused by the elastic response of the material. For small drops, the deformation of the interface remains mild and the corresponding shape is not hydrodynamically favourable, thus the terminal velocity remains small.

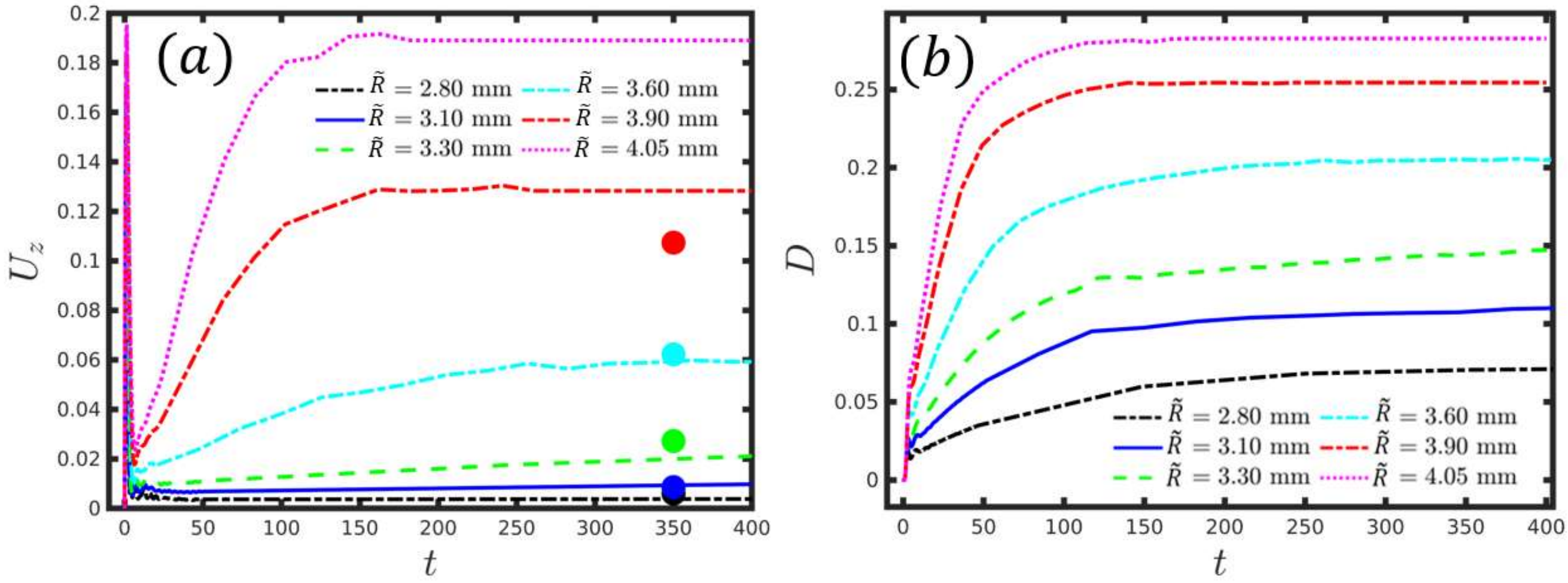

**Fig. 14:** Time evolution of the sedimentation velocity (a) and Taylor deformation parameter $D$ (b) for drops of different radii sedimenting in a solution of Carbopol 0.07%.

In **Fig. 15** we report a map of the evolution of drop shapes for different effective radii, together with the corresponding yield surfaces. For short times ($t$ = 10), the initial motion of the drop yields the surrounding YSM and the corresponding yield surface is almost symmetric, since the viscoelastic stresses have not developed yet. Proceeding in time, the asymmetry in the yield surface becomes more and more pronounced as the size of the drop increases, indicating the building up of viscoelastic stresses in the yielded region surrounding the sedimenting drop. The fact that the elastic response of the YSM is triggered by bigger drops is something that differs from what has been observed with bubbles rising in Carbopol, where small bubbles acquire an inverted teardrop shape and bigger bubbles tend to assume an oblate shape due to the more prominent inertial forces. The different behaviour between bubbles and drops has been documented also in polymeric viscoelastic fluids, where viscous drops are more prone to develop long and thick tails due to the lower values of the interfacial tension [52].

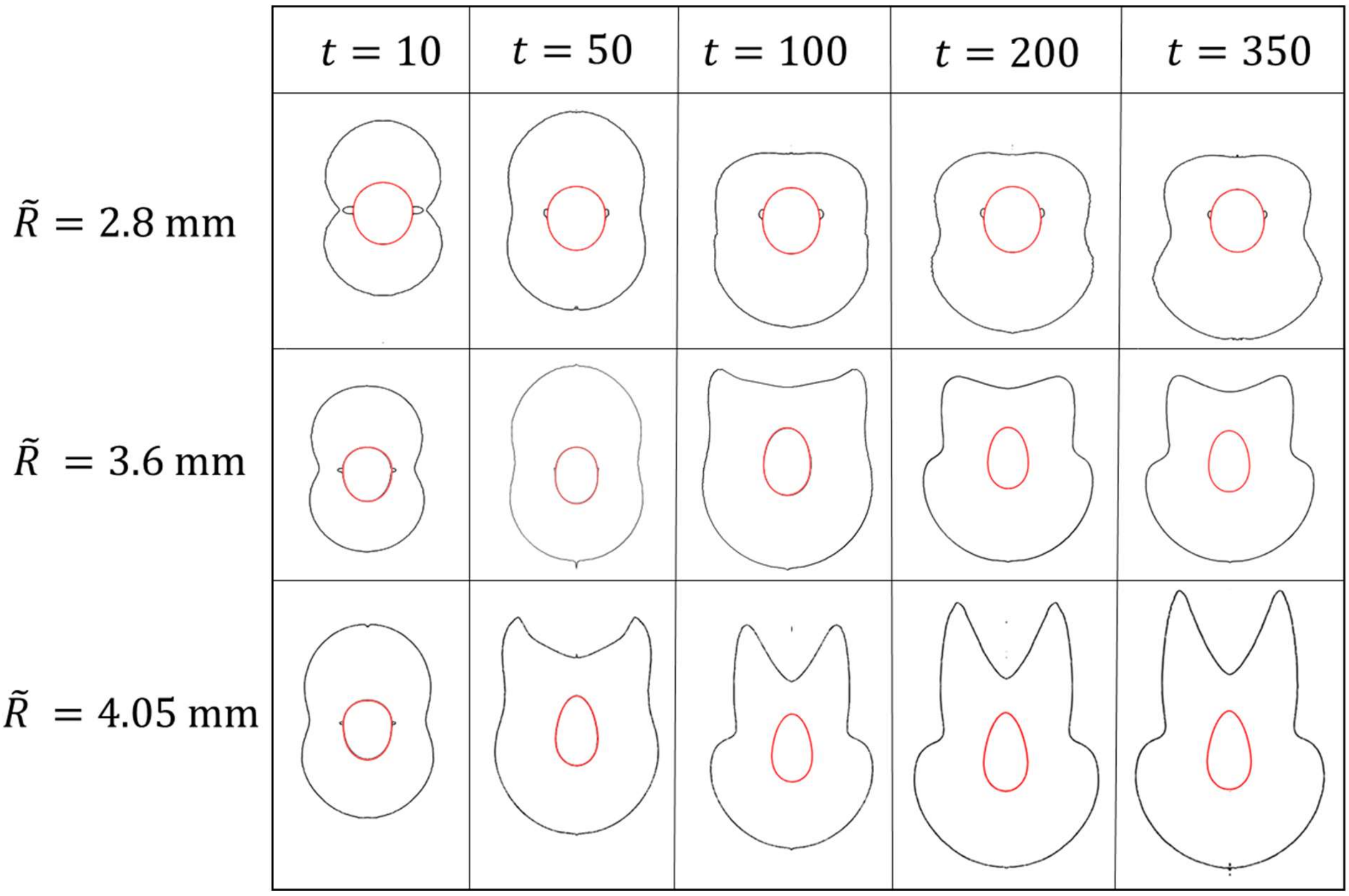

**Fig. 15:** Map of drop shapes (red) and yield surfaces (black) for different dimensionless times and radii

**Table V:** Comparison of the terminal velocities predicted in the present study with the corresponding steady velocities reported in [35]. The values without parenthesis are dimensionless, the ones in parenthesis have units of cm/s.

| $\tilde{R}$ (mm) | $u_z\|_{simulations}$ | $u_z\|_{experiments}$ |
|---|---|---|
| 2.80 | 0.0044 (0.07) | 0.006 (0.1) |
| 3.10 | 0.0126 (0.22) | 0.008 (0.15) |
| 3.30 | 0.022 (0.40) | 0.027 (0.49) |
| 3.60 | 0.0581 (1.09) | 0.062 (1.17) |
| 3.90 | 0.1283 (2.51) | 0.107 (2.10) |
| 4.05 | 0.1891 (3.77) | - |

### *3.3 Minimum radius for the entrapment of a viscous drop.*

In this subsection, we explore the entrapment conditions for a viscous drop sedimenting in YSMs. From a technological perspective, one of the most important characteristics of YSMs is their ability to entrap objects when the driving force which causes their motion is not able to overcome the yield stress. On the contrary, sedimentation always occurs in Newtonian or viscoelastic fluids. Hence, it is important to determine the conditions under which sedimentation is blocked by the plastic response of the continuous phase. Previous numerical studies concerning the rise of air bubbles in YSMs, employing (inelastic) viscoplastic constitutive equations [21] showed that the critical $Bn$ at which the bubble stops moving monotonically increases with $Bo$, i.e. the deformability of the interface promotes the motion. Furthermore, it has been shown that the entrapment condition of a solid sphere sedimenting in an EVP material also depends on the elastic response of the material, in particular the critical $Bn$ increases for more elastic materials [16]. The simultaneous effect of capillarity, elasticity and plasticity makes the entrapment conditions harder to determine. We decided to follow the experimental protocol reported in LHN [34] and decrease progressively the effective radius of the drop, fixing the material properties, with the goal to determine the radius of the minimum mobile drop, being aware that the corresponding critical conditions depend also on the Bond number (capillarity) and the Deborah number (elasticity). On the other hand, in §3.4 we explore the effect of changing solely the Bingham (yield stress) and the Bond numbers, to assess the independent impact of these parameters on the sedimentation dynamics.

Following previous experimental [23] and numerical [30] studies, we correlate the entrapment conditions to the terminal drag coefficient of the sedimenting drop, whose steady state value (when the drag force is balanced by buoyancy) solely depends on the terminal velocity:

$$C_d = \frac{8}{3U_t^2} \tag{28}$$

Although Eq. (28) holds true only for spherical objects, in the proximity of the entrapment condition the shape of the drop is almost spherical, thus we expect such approximation to be acceptable.

It is important to underline that the SHB model describes the YSM in the unyielded regions as a viscoelastic solid. This implies that the unyielded material is allowed to experience elastic deformations, hence, the motion of the drop is hindered but not completely nullified. Consequently, as the drop approaches entrapment conditions, we anticipate that, at long times, the lines of the travelled distance against time will tend to become horizontal, although their slope will never reach zero. Therefore, it is necessary to define a threshold in terms of terminal velocity or, conversely, in terms of drag coefficient, corresponding to the entrapment condition. This threshold is set at $C_d = 10^6$, corresponding to a terminal dimensionless velocity of $u_z \approx 0.00163$, following the experimental study of LHN [34]. These authors also report a power function that correlates the yield stress of the material with the critical radius below which the drop is immobile, but no indications are provided concerning the observation time and the minimum distance detectable. For the 0.07% Carbopol solution, the critical radius is reported to be equal to $\tilde{R} = 2.5$ mm, a result which is in quantitative agreement with our predictions reported in **Fig. 16**.

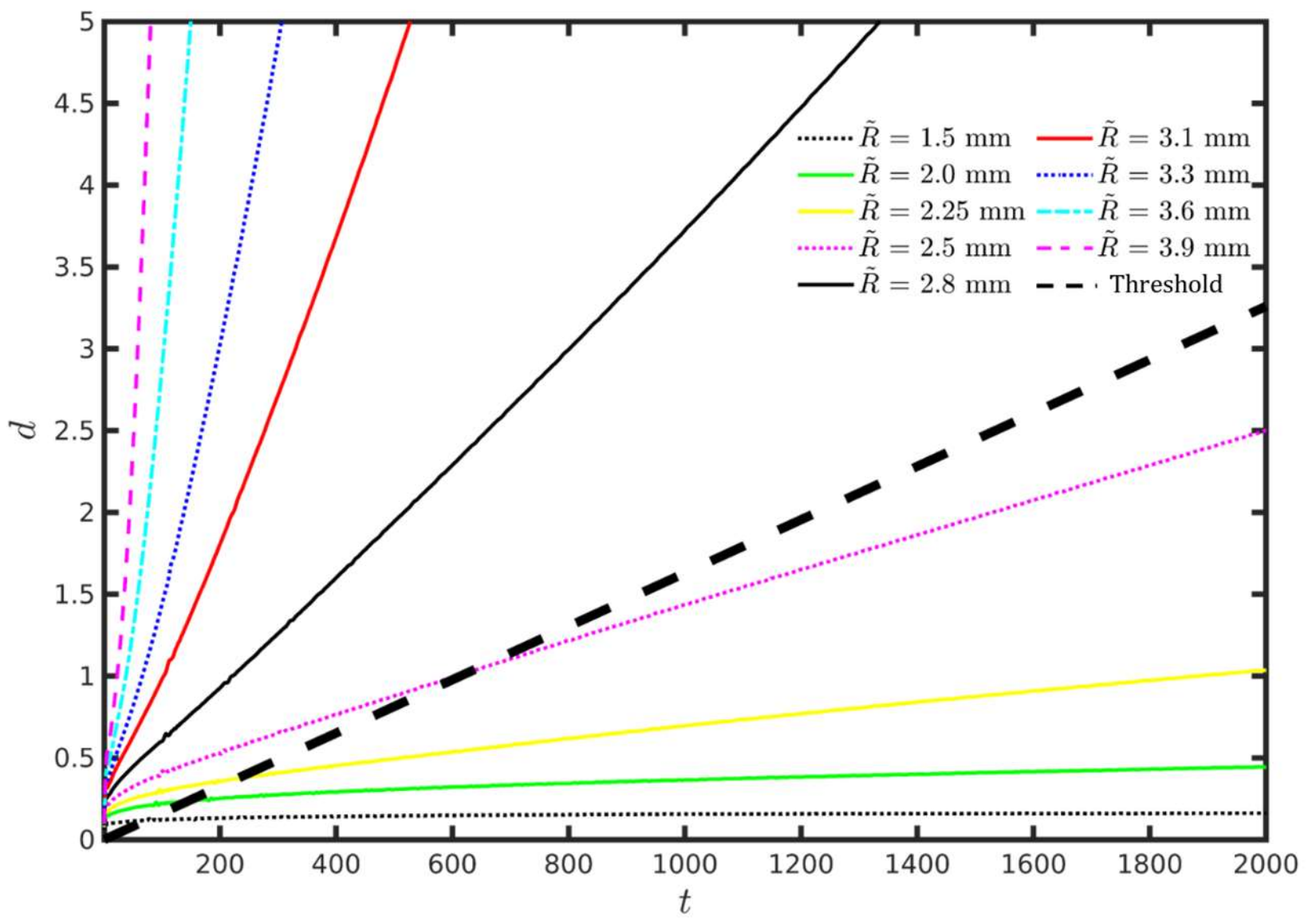

**Fig. 16:** Effect of the radius of the equivalent spherical drop $(\tilde{R})$ on the dimensionless distance travelled, $d$, as a function of the dimensionless time, $t$. The black/dashed line corresponds to the distance travelled by an object having a steady dimensionless velocity of $u_z \approx 0.00163$, corresponding to a drag coefficient of $C_d = 10^6$.

Here, we show the time evolution of the distance travelled by the sedimenting drop for several effective radii. The black/dashed line indicates the distance travelled by a steady object having a velocity of $u_z = 0.0016$, corresponding to the threshold drag coefficient of $C_d = 10^6$. We observe that a reduction in the equivalent radius of the drop monotonically decreases the slope of the curves. Please note, the infinitesimal distance covered by the drop of radius $\tilde{R} = 1.5$ mm after an exceedingly long time. The very low velocity could remind us of the pitch drop experiment in Queensland [54]. Since this slope represents the velocity of the drop, the approach to a horizontal asymptote indicates the stasis of the object, i.e., the entrapment condition. As a consequence, all the curves above the black dashed line represent mobile drops, while the curves below indicate entrapment.

In **Fig. 17**, we report the dependence of the terminal drag coefficient on the Bingham number. The black dashed curve is obtained by varying the effective radius of the drop and tracking the terminal drag coefficient as a function of the Bingham number. We remark that a change in the size of the drop has an influence on all the dimensionless numbers, affecting the interplay of elastic, capillary and plastic forces. For example, a change in the interfacial tension of the two fluids would modify the depicted curves. Indeed, as documented also for inelastic viscoplastic materials, the entrapment condition strongly depends on the deformability of the objects [21]. In **Fig. 17**, the entrapment condition is represented by a vertical asymptote of the $C_d$ in correspondence of a critical Bingham number. The critical Bingham number found in this study (around 0.10) is intermediate between the reported values for solid spheres [15] and air bubbles [21], something which is consistent with our physical intuition. We repeat the simulations for a lower elastic modulus, fixing the other material parameters, hence increasing the Elastogravity number (modulating the elastic response of the material) and report the results with the cyan dashed line. We observe an increase in the critical Bingham number when the elastic response of the material is enhanced, mainly for two reasons. Primarily, a lower $G$ allows a higher deformation of the EVP material prior to yielding, promoting the motion of the deformable intrusion. On the other hand, the development of higher viscoelastic stresses in the proximity of the interface enhances its deformation, which has been proven to be beneficial for the translational motion [30]. We remark that such dependence of the entrapment conditions on the elastic response of the material is consistent with previous observations concerning air bubbles rising [30] and solid spheres sedimenting in YSM [16]. The straight vertical lines in **Fig. 17** indicate the critical $Bn$ reported in previous numerical studies respectively for solid spheres (red-dotted, [15]) and air bubbles (blue-dashed-dotted, [21]) moving in inelastic viscoplastic materials, and for air bubbles rising in elastoviscoplastic fluids (pink-dashed, [30]).

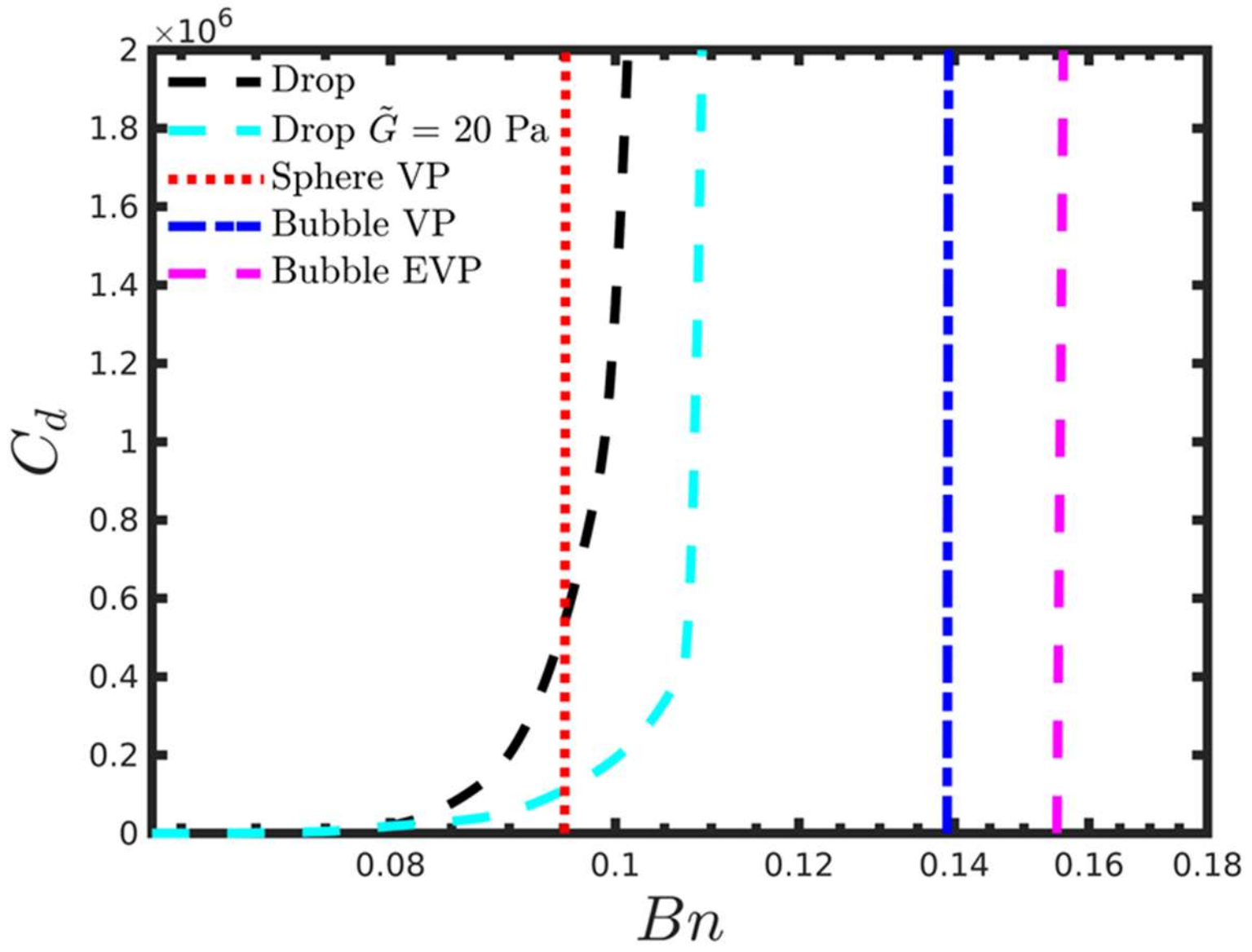

**Fig. 17:** Terminal drag coefficient as a function of the Bingham number. The black/dashed line indicates present results for viscous drops sedimenting in a 0.07% solution of Carbopol, the cyan/dashed line indicates present results for the same drops sedimenting in a Carbopol solution with lower elastic modulus (higher $Eg$) with the same remaining dimensionless numbers. The vertical lines indicate the critical Bingham number reported in the literature for (inelastic) viscoplastic materials for solid spheres (red dotted line) [15], for air bubbles (blue / dash-dotted) [21] and air bubbles in elastoviscoplastic materials (pink dashed line) [30].

### *3.4 Parametric study: effect of the rheological parameters on the sedimentation dynamics*

In this subsection, we analyze the effect of the rheological parameters on the dynamics of the sedimenting drop. Since, as previously mentioned, all the dimensionless numbers depend on the value of the effective radius $\tilde{R}$, we select a base case at $\tilde{R} = 3.6$ mm and vary independently the values of the yield stress $\tilde{\sigma}_y$ (which solely affects $Bn$), the elastic modulus $\tilde{G}$ (which only affects $Eg$), the interfacial tension $\tilde{\Gamma}$ (corresponding to a variation of $Bo$). The dimensionless numbers corresponding to the base case are $Ar = 1.46,\ Bo = 1.79,\ Bn = 0.068,\ Eg = 0.69,\ \rho° = 1.621,$ $\eta° = 0.003,\ \beta = 0.1$.

#### *3.4.1 Effect of the yield stress (Bingham number)*

When inelastic viscoplastic materials are considered, an increase in $Bn$ monotonically decreases the terminal velocity of a rising bubble [21], [22] or a sedimenting drop [37]. On the other hand, the analysis of the transient velocities and deformation parameter $D$, reported in **Fig. 18**, suggests something different. Starting from a very small value of $Bn$, we progressively increase its value up to the critical $Bn$ found in §3.3. All the curves manifest an initial overshoot (i.e. acceleration period) followed by a decrease in the velocity at very short time scales (leftmost part of the **Fig. 18** (a), up to $t \approx 10$). Subsequently, we observe an acceleration period whose duration increases with $Bn$. Such acceleration period coexists with the development of the teardrop shape, as reported in the right panel of **Fig. 18**.

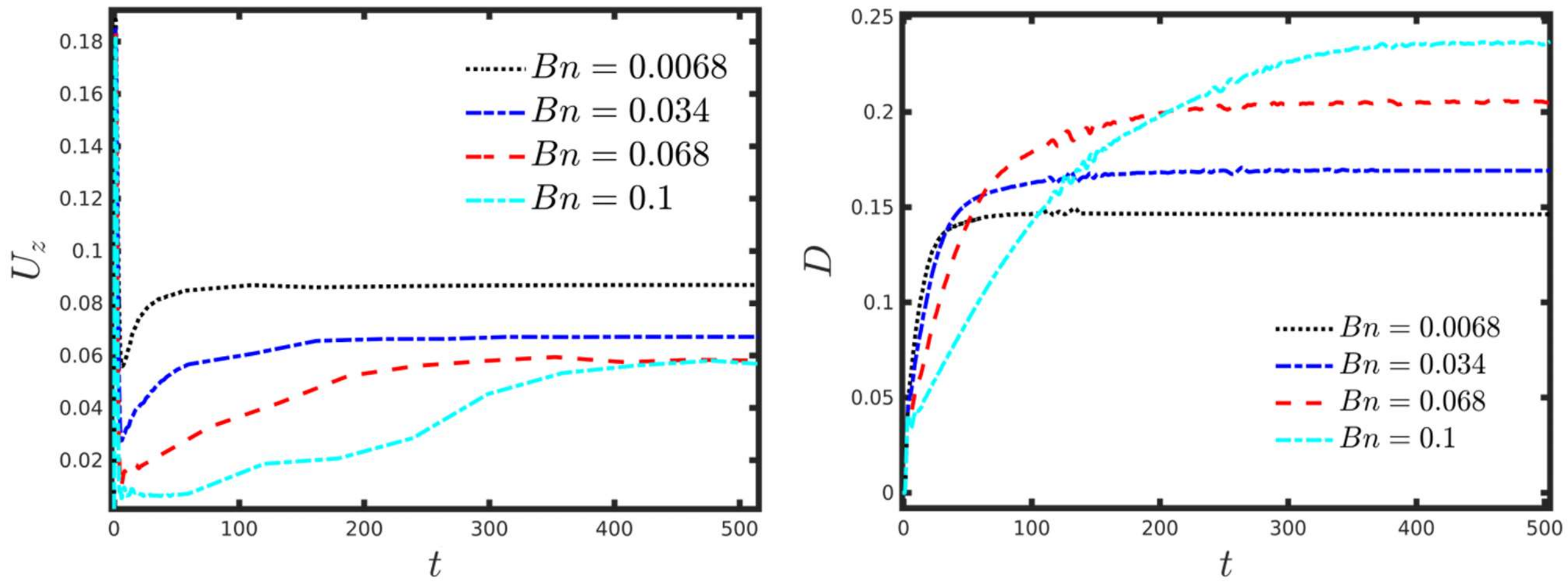

**Fig. 18:** Transient development of the velocity and the Taylor deformation parameter $D$ for a sedimenting drop having $\tilde{R} = 3.6$ mm at different $Bn$ (i.e., yield stress $\tilde{\sigma}_y$).

For high $Bn$, very long times (up to 2,000) are required to develop the terminal shape and velocity, see **Fig. 19**. We report the values of the terminal sedimentation velocity plotted against the $Bn$ number in **Fig. 20**. Consistently with the transient profiles reported in **Fig. 18**, the increase of $Bn$ involves an initial reduction of the terminal velocity caused by the higher material plasticity. Nonetheless, for higher $Bn$, the terminal sedimentation velocities approach an almost constant value, indicating that the increase of resistance caused by the higher material plasticity is balanced by the stronger elastic response in the fluidized part of the material, which allows the drop to acquire a more hydrodynamic shape and enhance its motion.

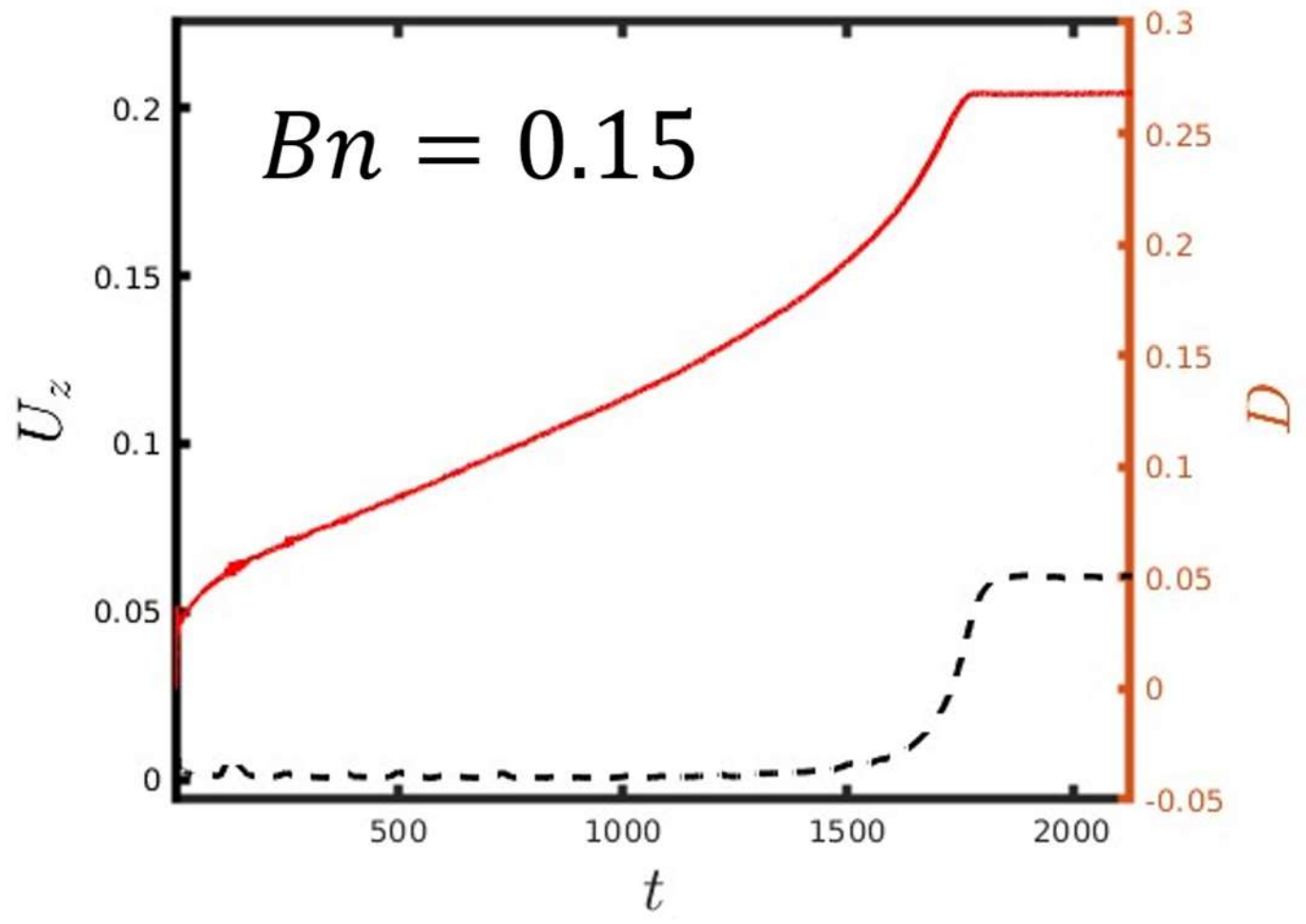

**Fig. 19:** Transient development of the velocity and the Taylor deformation parameter $D$ for a sedimenting drop having $\tilde{R}$ = 3.6 mm at $Bn$ = 0.15.

Eventually, for even higher *Bn,* the buoyancy force is not able to overcome the yield stress of the material and the drop cannot adequately deform, hence the Taylor parameter is drastically reduced and the entrapment condition is approached, see **Fig. 20**.

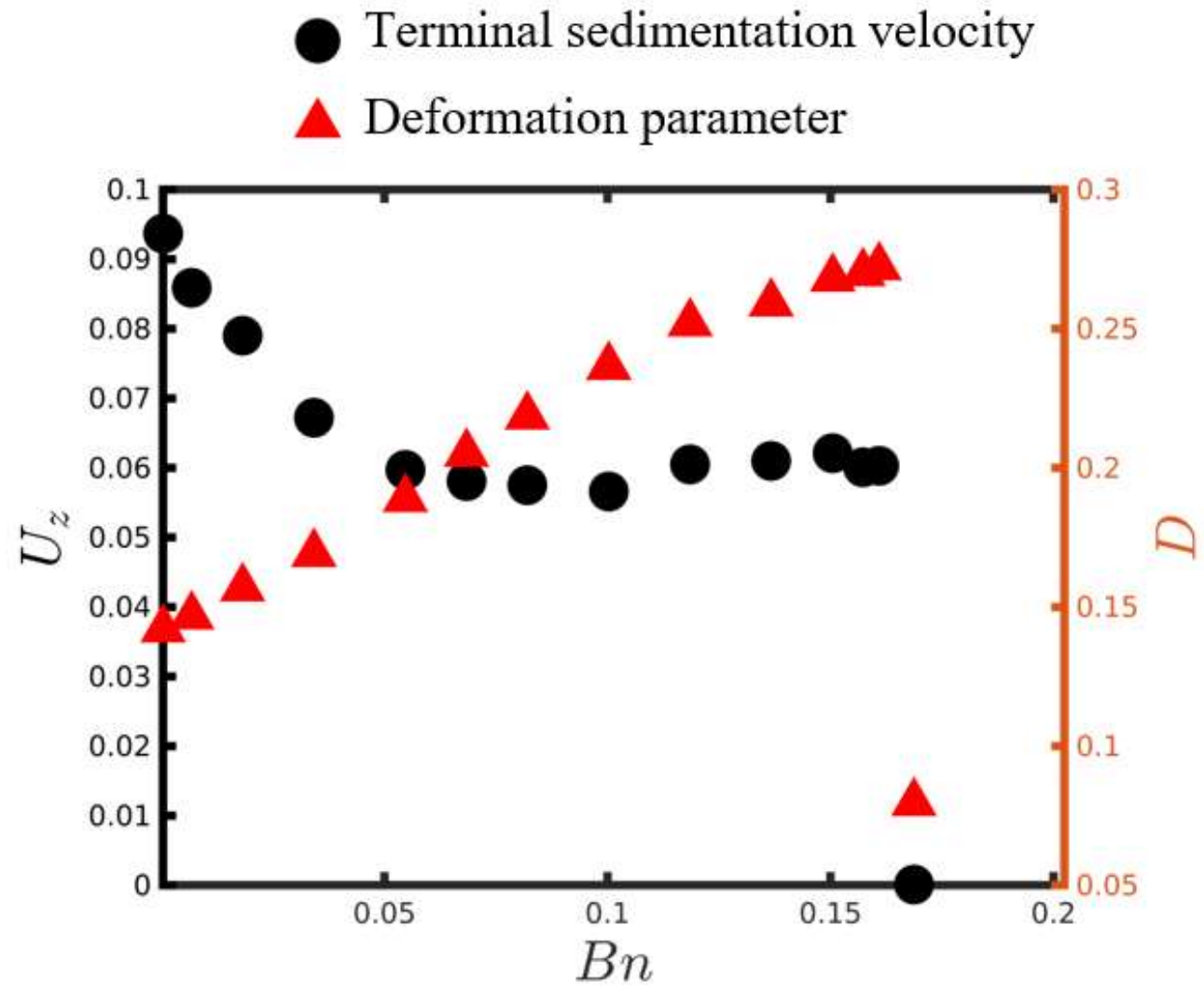

**Fig. 20:** Terminal sedimentation velocity and deformation parameter as a function of the $Bn$ number for a viscous drop of radius $\tilde{R}$ = 3.6 mm.

Motivated by the long times required to reach a steady motion at high $Bn$ (**Fig. 20**), we waited up to $t$ = 10,000 for the case at $Bn$ = 0.17, still observing no increase in the transient velocity.

It is noteworthy that the critical $Bn$ above which the drop stops moving is much higher than the one obtained in §3.3 and is very close to the one reported for air bubbles rising in EVP materials at similar Bond numbers [30]. This observation presents further evidence that the entrapment conditions are strongly affected by the deformability and the shape of the sedimenting drop.

When $Bn$ increases, we observe a shrinkage of the fluidized region, because velocity variations are limited closer to the sedimenting drop. For this reason, velocity gradients increase locally and lead to the increase of viscoelastic stresses, triggering a stronger elastic response of the fluidized material, as indicated in **Fig. 21**, where we compare the viscoelastic stress distribution around a sedimenting drop of effective radius $\tilde{R} = 3.6$ mm for $Bn = 0.038$ and $Bn = 0.15$. These higher stresses are responsible for the development of the teardrop shape and for the increase of the sedimentation velocity.

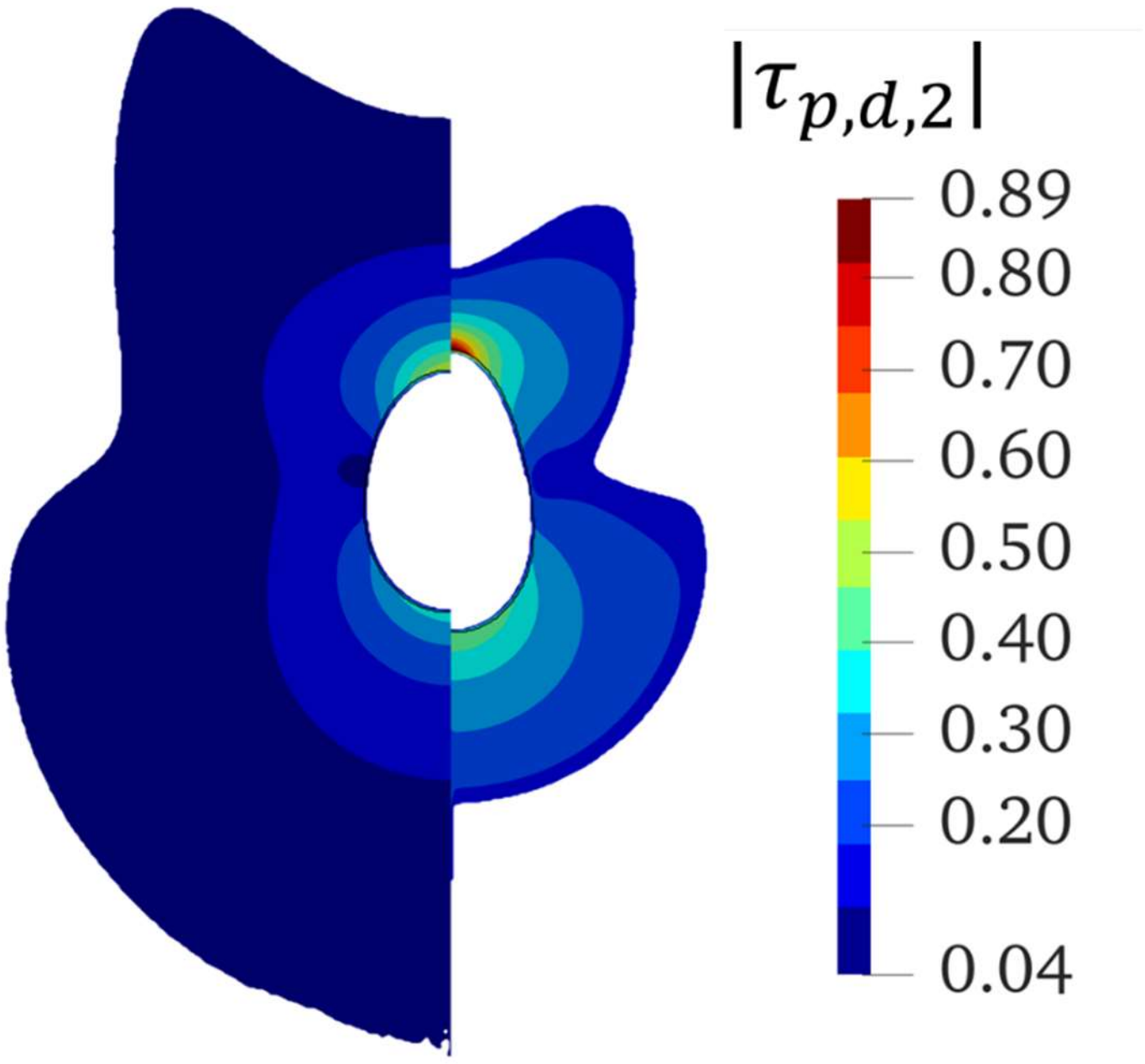

**Fig. 21:** Magnitude of the viscoelastic stresses, defined in **Eq. (5)**, in the yielded regions, around a sedimenting drop of radius $\tilde{R} = 3.6$ mm for $Bn = 0.038$ (left) and $Bn = 0.15$ (right).

### 3.4.2 *Effect of the elastic modulus (Elastogravity number)*

The elastic modulus $\tilde{G}$ is a measure of the elasticity of the material and is inversely proportional to the relaxation time of the YSM. Fixing all the other dimensionless numbers, an increase in $\tilde{G}$ results in a reduction in the Elastogravity number, hence the elastic response becomes less prominent. The asymptote of $\tilde{G} \to \infty$ reduces the SHB constitutive equation to the well-known Herschel-Bulkley model, where elastic effects are absent, and the continuous phase behaves as an ideal shear-thinning viscoplastic material. Nonetheless, the stress in the unyielded regions remains uniquely determined and the material there behaves as a viscoelastic-solid. With the goal to assess the effect of $\tilde{G}$ on the flow conditions, we report the Taylor deformation parameter $D$ and the transient dimensionless velocity $U_z$ as a function of the dimensionless time $t$.

In **Fig. 22**, we report the transient development of the sedimentation velocity of the drop, together with the Taylor deformation parameter $D$. The trend is monotonic and indicates that at higher $Eg$, both the terminal velocity and the terminal deformation parameter $D$ are higher, meaning that the sedimentation is faster, and the shape is more prolate. Such observation provides evidence of the crucial role played by the elastic response of the material for the development of the teardrop shape, something that is not observed for bubbles and drops moving in inelastic viscoplastic materials. All the cases, regardless of the values of $Eg$, show an initial acceleration which is primarily governed by the buoyancy force, opposed by the viscous resistance and the plasticity of the YSM, followed by an overshoot indicating the development of the elastic stresses, as already observed for drops rising in viscoelastic fluids [52]. We observe a clear distinction between the dynamics of drops sedimenting in highly elastic (high $Eg$, black and blue lines in **Fig. 22**) and mildly elastic YSM (low $Eg$, red, cyan and green lines in **Fig. 22**). In the former cases, the elastic stresses are sufficiently high to promote the development of a prolate shape, thus, the drop experiences a second acceleration and reaches a terminal velocity which increases with $Eg$, while in the latter ones such acceleration is not observed, and the terminal velocity is much lower. We report the terminal shapes for all the $Eg$ explored in **Fig. 22**. In the viscoplastic limit the terminal shape deviates much less from the initially spherical one and the sedimenting drop assumes a slightly prolate conformation, similar to what is observed for bubbles rising in HB materials [22].

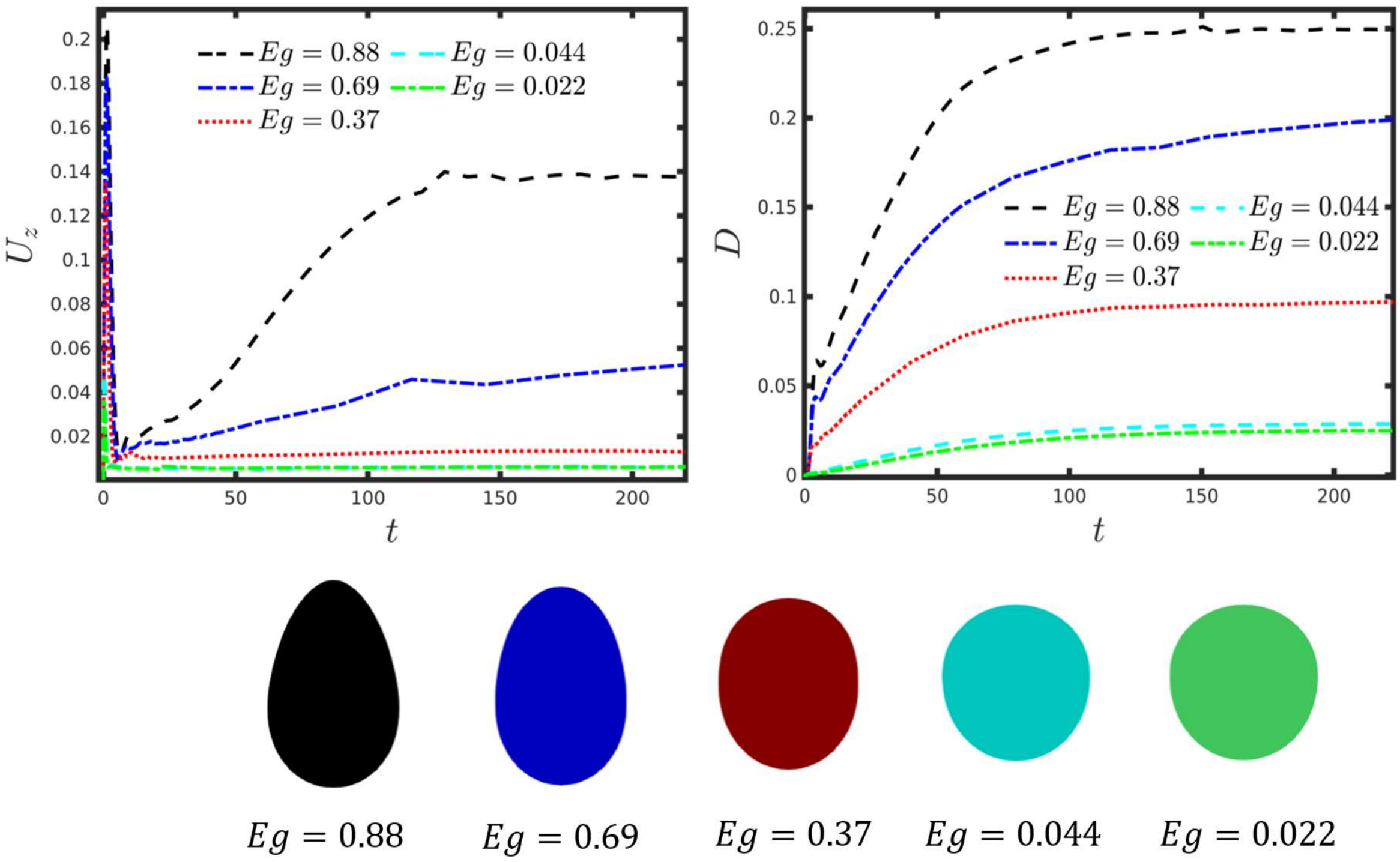

**Fig. 22**: Transient development of the velocity and the Taylor deformation parameter D for a sedimenting drop having $\tilde{R} = 3.6$ mm at five values of $Eg$ (top). Terminal drop shapes for the corresponding $Eg$ (bottom).

Furthermore, it is instructive to analyze the distribution of the yielded regions around the sedimenting drop in two cases at $Eg = 0.88$ ($\tilde{G} = 25$ Pa) and $Eg = 0.37$($\tilde{G} = 60$ Pa). The extent and the shape of such areas are strongly affected by the elastic modulus $\tilde{G}$, as observed in **Fig. 23**. Indeed, when the elastic response of the material is enhanced, the fore-and-aft asymmetry is more pronounced, indicating that stronger elastic stresses are developed around the sedimenting object. Lower values of the elastic modulus $\tilde{G}$ correspond to higher relaxation times, thus the viscoelastic stresses accumulate for longer times, intensifying their magnitude, and relax further away from the sedimenting drop. Such stresses contribute to the deformation of the interface between the two fluids, allowing the drop to acquire a more hydrodynamic shape with a consequent reduction in the drag and an increase in the terminal velocity. It is interesting that although the terminal velocity of the drop is higher for $Eg = 0.88$, the extent of the yielded region in front of it is higher for $Eg = 0.37$. This observation is consistent with what has been observed for bubbles [30] and rigid spheroids [55] moving in EVP materials, and can be justified by exploiting an argument presented in [56]: the maximum strain that a material can undergo before yielding is measured by the auxiliary dimensionless number $\varepsilon = \frac{\tilde{\tau}_y}{\tilde{G}} = EgBn$, namely the *yield strain*. Increasing the elastic modulus $\tilde{G}$ reduces $Eg$ and $\varepsilon$, when the yield stress of the material is kept constant. This implies that the material having higher $\tilde{G}$ can sustain smaller strains before yielding, and since the development of the stresses is mainly caused by the passage of the sedimenting drop, the material in front of it yields further in front of it with respect to its more elastic counterpart.

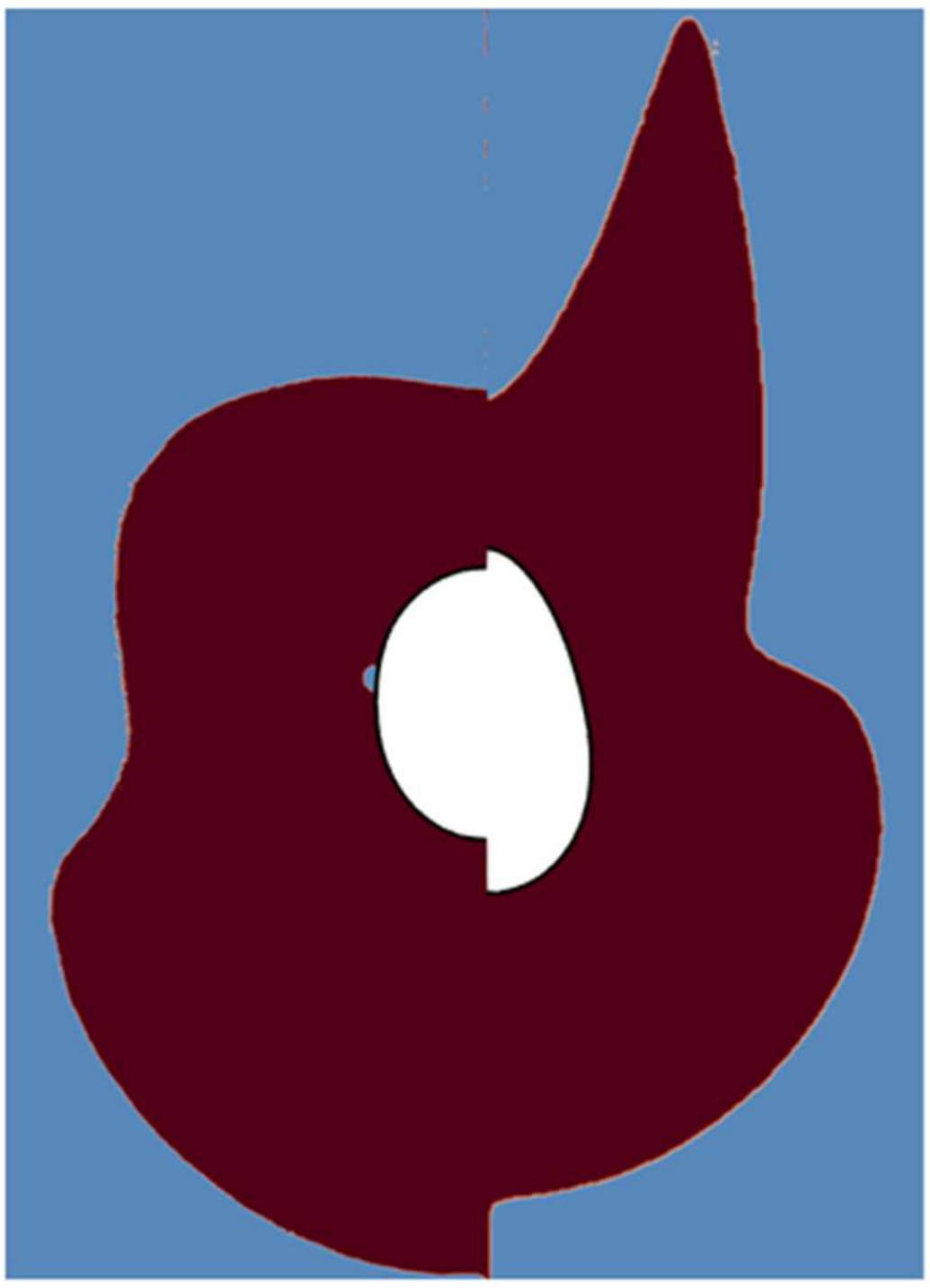

**Fig. 23**: Predicted terminal shapes and distribution of the yielded (red)/unyielded (blue) areas for a viscous drop of radius $\tilde{R} = 3.6$ mm sedimenting in a YSM. On the left $Eg = 0.37$, on the right $Eg = 0.88$.

### 3.4.3 *Effect of the interfacial tension (Bond number)*

The interfacial tension between the two fluids modulates the deformability of the viscous drop during its sedimentation. Depending on $\tilde{\Gamma}$ (i.e., on Bond number), the magnitude of the elastic stresses may or may not suffice to actually change the shape of the liquid object, triggering an increase in the terminal velocity. Previous studies concerning the buoyancy driven motion of drops in Bingham-like materials [37] report a moderate increase in the terminal velocity for increasing capillary numbers, which are proportional to $Bo$.

However, it is important to underline that in such studies the elastic response of the material is not considered, meaning that the moderate deformation experienced by the drops is solely due to the interplay between inertial, viscous, and plastic effects. On the other hand, in the present study the drop is prone to attain a teardrop shape with a consequent substantial increase in the terminal velocity. We vary progressively the value of the interfacial tension $\tilde{\Gamma}$ in the span $\tilde{\Gamma} \in [0.037 - 0.1]$ N/m, since the vast majority of liquid-liquid systems are reported to lie in such range. The corresponding Bond numbers are reported in **Table VI**.

**Table VI:** Values of the interfacial tension $\tilde{\Gamma}$ and corresponding Bond numbers.

| $\tilde{\Gamma}$ (N/m) | *Bond number* |
|---|---|
| 0.037 | 2.1 |
| 0.044 | 1.80 |
| 0.06 | 1.3 |
| 0.1 | 0.79 |

As expected, the drop experiences a stronger deformation at higher Bond numbers, with the Taylor parameter D reaching higher and higher terminal values, see **Fig. 24**. Similarly to what we reported in §3.4.2 for lower *Eg*, a stronger deformation facilitates the motion of the sedimenting drop, with a consequent increase in the terminal velocity.

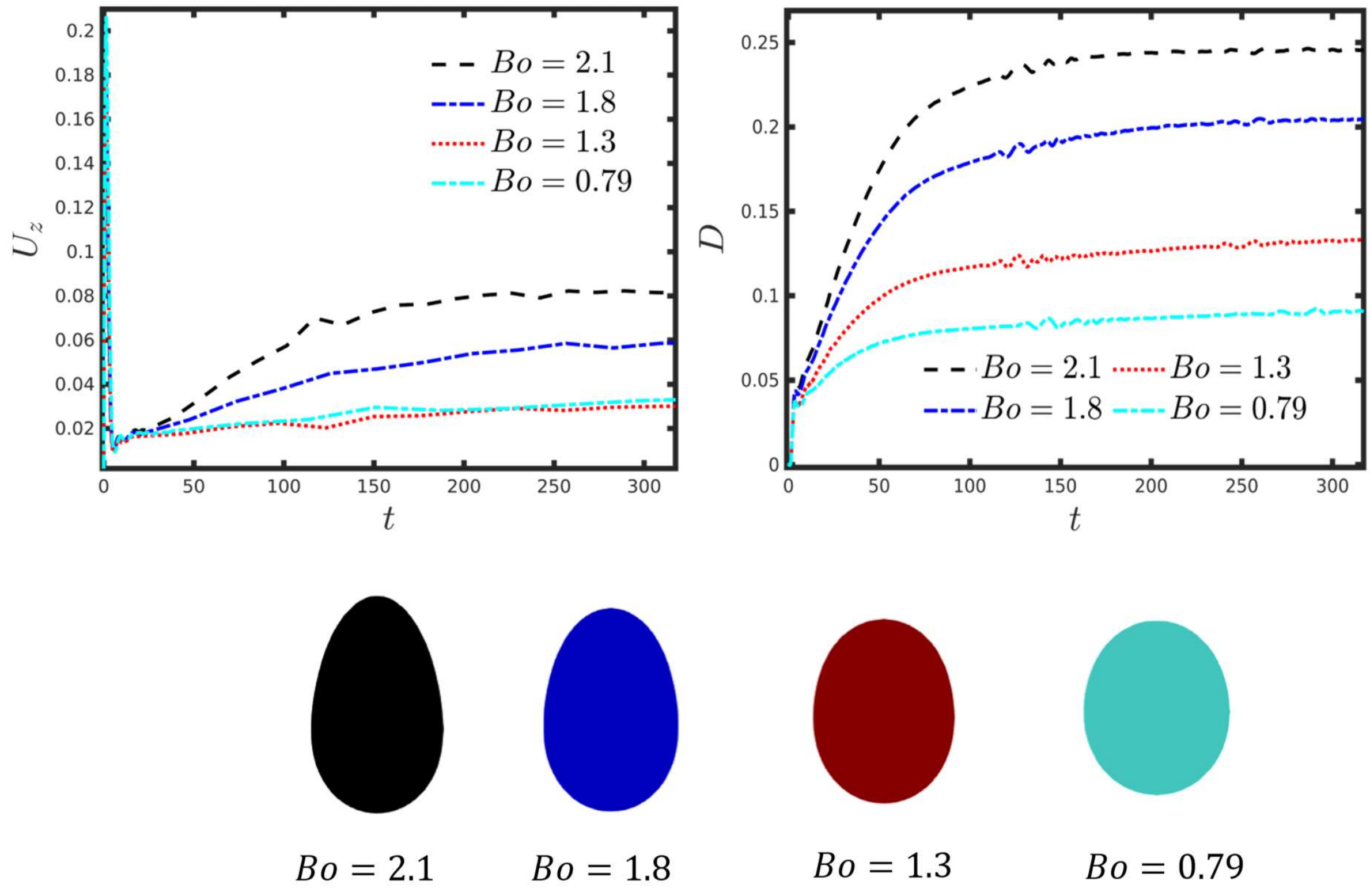

**Fig. 24:** Time evolution of the sedimentation velocity and deformation parameter of a drop having $\tilde{R} = 3.6$ mm for different values of the interfacial tension. On the bottom of the figure, the terminal shapes obtained for four different Bond numbers.

## 4. Final remarks and conclusions

In this work, we present the first numerical study concerning the sedimentation of a viscous drop in a yield stress material exhibiting elastic effects. The rheology of the continuous phase is described by the Saramito-Herschel-Bulkley constitutive equation. Initially, we place the viscous drop in the proximity of the upper boundary of a cylindrical container and investigate the transient development of the velocity profile, the shape of the interface and the yield surface, i.e., the separatrix between solid-like and liquid-like regions in the continuous phase. We validate our numerical setup via comparison with previous numerical [30] and experimental [23] results concerning the rising of an air bubble in a yield stress material (YSM). Then, we replicate the experimental findings of Lavrenteva-Holenberg-Nir (LHN) [35]. The terminal velocities are in quantitative agreement with the experimentally reported ones and the mechanism behind the generation of the characteristic teardrop shape is elucidated via the analysis of the viscoelastic stress field around the sedimenting drop. Elastic effects are found to be crucial to justify the deviation from the theoretical results obtained for (inelastic) viscoplastic materials. In particular, it is necessary to include the elastic properties to correctly predict phenomena that have been reported in experiments, i.e., the development of a teardrop shape and the inversion of the flow field behind the sedimenting object (negative wake).

We investigate the entrapment conditions following the experimental protocol, thus reducing progressively the radius of the viscous drop and monitoring the terminal velocity and the drag coefficient. The predicted minimum radius for starting the motion is in quantitative agreement with the experimentally reported one. Subsequently, we assess the effect of the rheological parameters through an extensive parametric study, where we vary independently the yield stress, $\tilde{\sigma}_y$, the elastic modulus, $\tilde{G}$ and the interfacial tension, $\tilde{\Gamma}$. The complex interplay between plastic and elastic forces is explored, observing that an increase in the material plasticity first reduces the terminal sedimentation velocity, while at higher $Bn$ the elastoplastic response of the material promotes the deformation of the drop and balances the increasing resistance, stabilizing the terminal sedimentation velocity. More deformable drops acquire higher terminal velocities due to the development of the teardrop shape similar to the one observed for air bubbles rising in YSM. Such shape is characterized by a reduced drag that favors its mobility and retards its entrapment.

We plan to extend this study considering the motion of two viscous drops coaxially sedimenting in a YSM, inspired by previous experimental [34]-[36] and theoretical works [57] with the scope to elucidate the physical mechanism behind their interaction when elastic and plastic effects are both taken into account. Furthermore, it would be interesting to extend our numerical setup toward a fully 3D simulation, to analyze possible asymmetries in the flow-field and/or drop deformation, as observed for air bubbles rising in viscoelastic solutions [18] or for drops rising next to a solid wall or in the same horizontal plane.

## Acknowledgement

This work was funded by the European Union's Horizon 2020 research and innovation program under the Marie Skłodowska-Curie grant agreement No 955605; project YIELDGAP (https://yieldgap-itn.com). The authors would like to thank Athanasios Kordalis, Pantelis Moschopoulos, Outi Tammisola and Kazi Tassawar Iqbal for insightful discussions.

## Authors declarations

The authors have no conflict of interest to declare.

## APPENDIX: CONVERGENCE TESTS

To ensure that the initial distance from the upper boundary ($d_0$) and the axial length of the domain ($L$) do not affect the dynamics of sedimentation, we perform a convergence test by varying such quantities. The results are shown in **Fig. A 1**, where a perfect superposition is obtained. In the bottom of the same figure, we also report the terminal shape of the drop for each case.

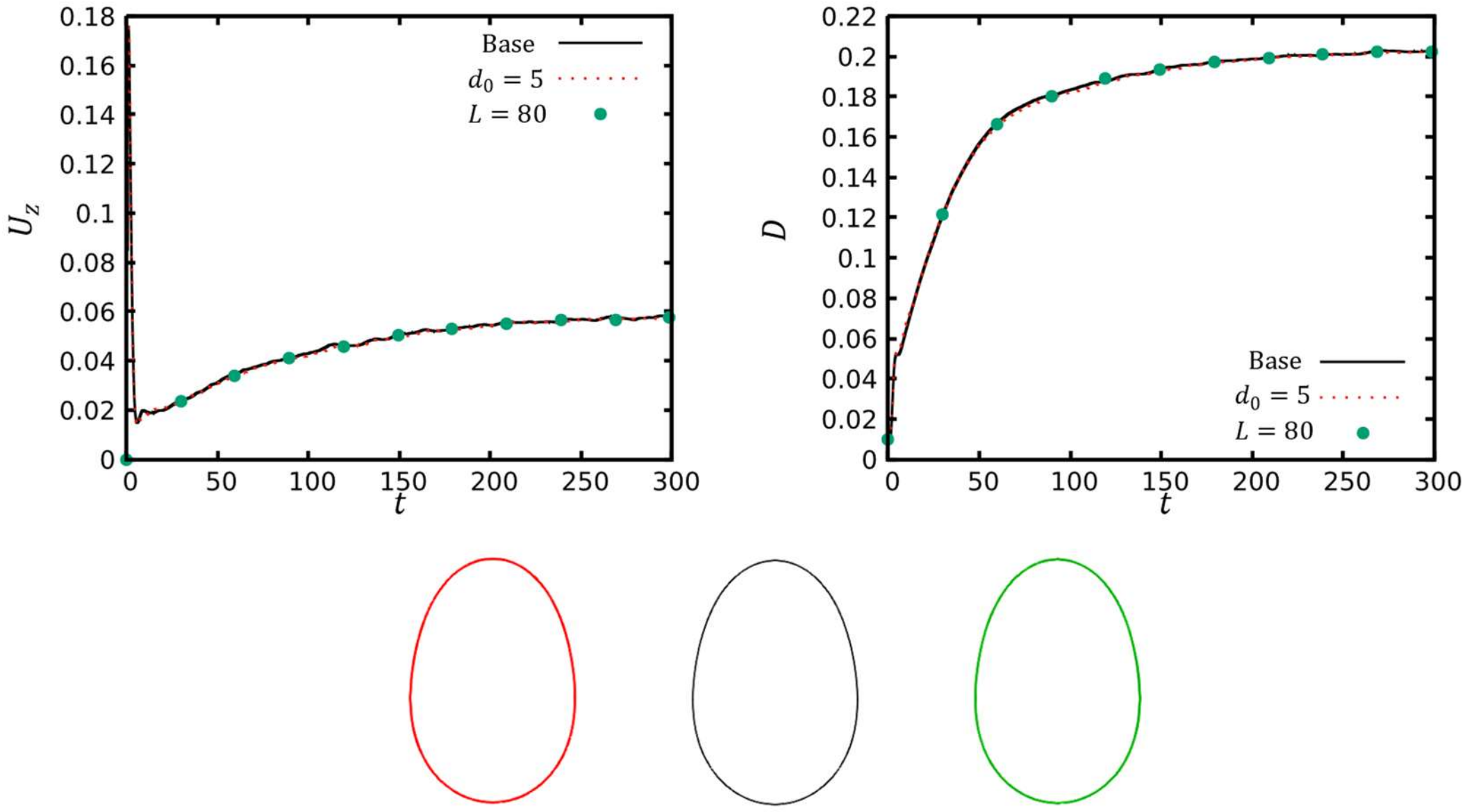

**Fig. A 1:** Time evolution of the sedimentation velocity and deformation parameter of a drop having $\tilde{R} = 3.6$ mm for a different initial distance from the upper container boundary or for a longer axial length of the domain than in the base case. On the bottom of the figure, the terminal shapes obtained at $t = 300$ for the three cases, from left to right: $d_0 = 5, L = 40$, base case with $d_0 = 10, L = 40$, and $d_0 = 10, L = 80$.

Furthermore, we perform a mesh convergence test to ensure that our results are independent on the resolution of the computational grid. We select three different values of the maximum level of refinement, namely $N_{max} = [11\,(\mathrm{M1}), 12\,(\mathrm{M2}), 13\,(\mathrm{M3})]$, and track both the transient velocity and deformation. The results are summarized in **Fig. A 2**.

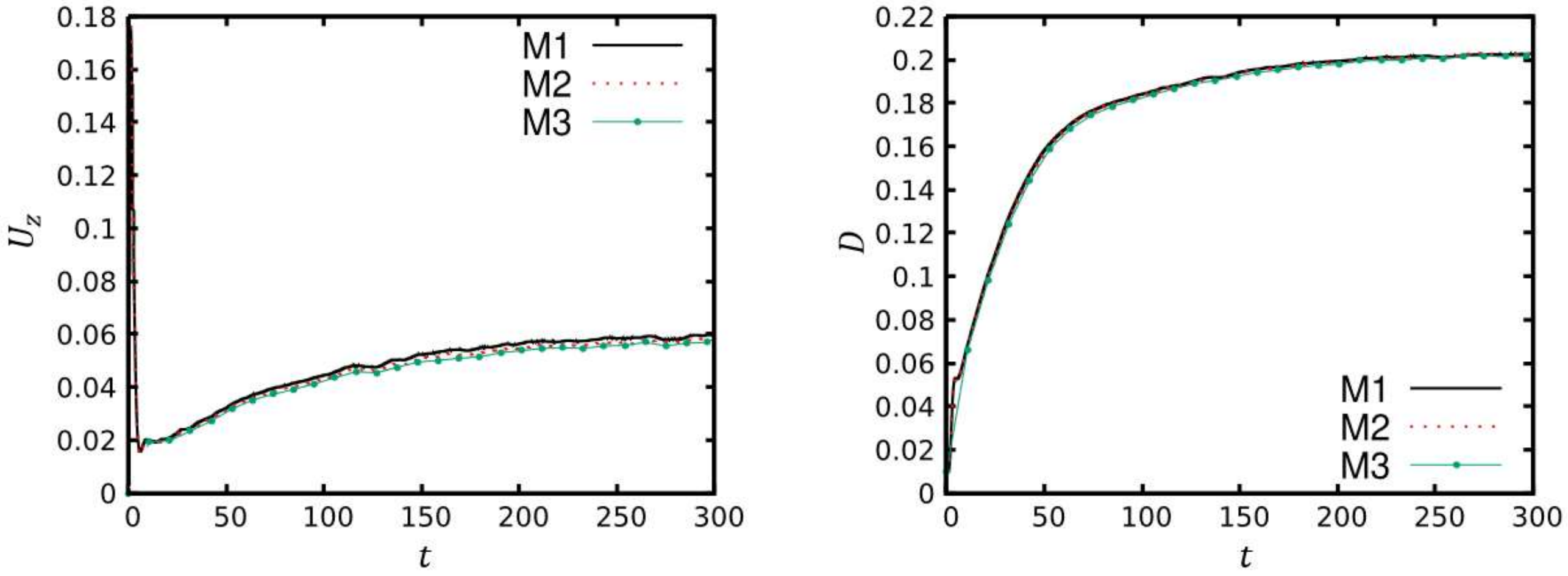

**Fig. A 2**: Time evolution of the sedimentation velocity and deformation parameter of a drop having $\tilde{R} = 3.6\,mm$ for three different computational grids.

The terminal values for the sedimentation velocity and deformation parameter are summarized in **Table A I**.

**Table A I:** Terminal sedimentation velocity and deformation parameter for three different computational grids and corresponding maximum level of refinement.

| Mesh | $N_{max}$ | cells/$R_{drop}$ | $u_z$ | $D$ |
|---|---|---|---|---|
| M1 | 11 | 51.2 | 0.0596 | 0.2028 |
| M2 | 12 | 102.4 | 0.0577 | 0.2020 |
| M3 | 13 | 204.8 | 0.0571 | 0.2017 |

Clearly, the choice of M2 is adequate and allows to obtain mesh-independent results.

We also perform a time-convergence study, whose results are displayed in **Fig. A 3** and summarized in **Table A III**, by varying the maximum time step admissible.

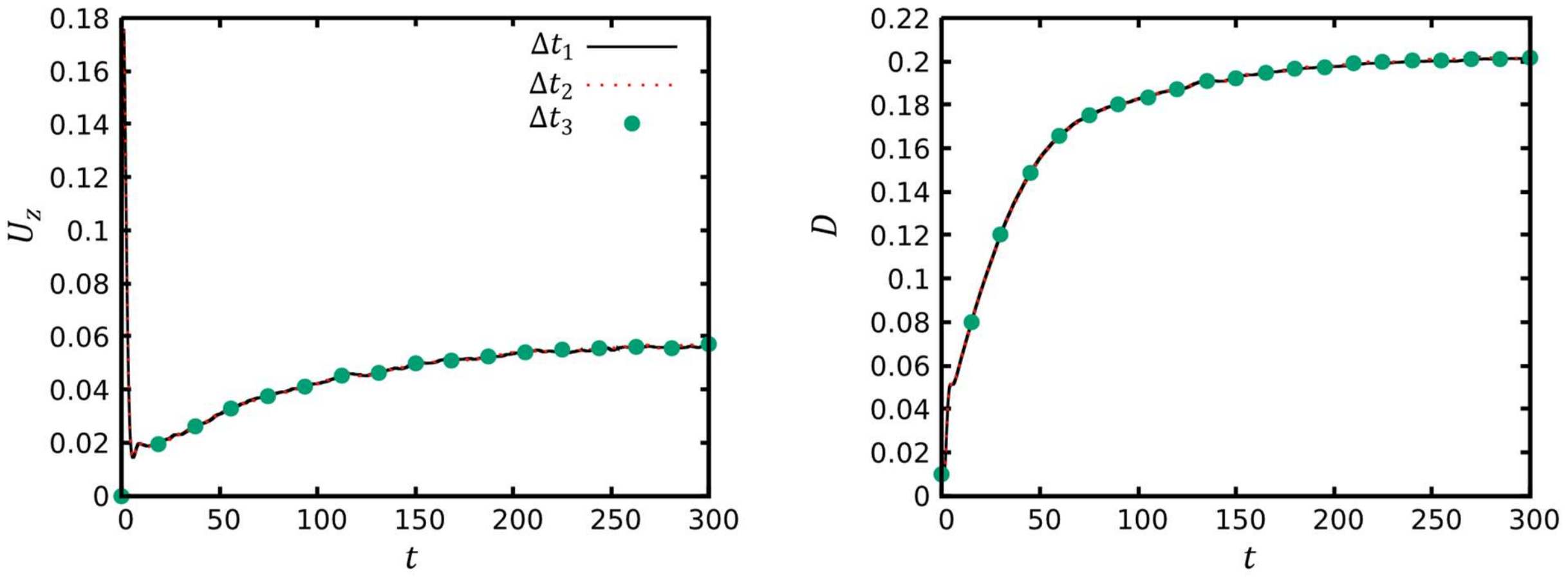

**Fig. A 3:** Time evolution of the sedimentation velocity and deformation parameter of a drop having $\tilde{R} = 3.6\, mm$ for three different maximum time steps.

The perfect superposition of the data indicates that the choice of $\Delta t_{max} = 7.5 \cdot 10^{-4}$ is adequate. This value respects the capillary time-scale condition introduced in the problem formulation, $\max(\Delta t) < \sqrt{\frac{\rho^{\circ}+1}{\rho^{\circ}-1}\frac{Bo\Delta x^3}{\pi}} \approx 1.5 \cdot 10^{-3}$.

**Table A III:** Terminal sedimentation velocity and deformation parameter for three different maximum time steps.

| Time step | $\Delta t_{max}$ | $u_z$ | $D$ |
|---|---|---|---|
| $\Delta t_1$ | $5 \cdot 10^{-4}$ | 0.0570 | 0.2016 |
| $\Delta t_2$ | $7.5 \cdot 10^{-4}$ | 0.0570 | 0.2016 |
| $\Delta t_3$ | $1 \cdot 10^{-3}$ | 0.0571 | 0. 2017 |